\documentclass[]{jfm}

\usepackage{graphicx}
\usepackage{epstopdf,epsfig}
\usepackage{newtxtext}
\usepackage{newtxmath}
\usepackage{natbib}
\usepackage{hyperref}
\usepackage{verbatim}
\usepackage{amsmath,amssymb,bm}
\usepackage{setspace}
\usepackage{caption}
\usepackage{subcaption}
\usepackage{breqn}
\usepackage{cleveref}
\usepackage[font=normal,labelfont=bf,
   justification=justified,
   format=plain]{caption} 
\hypersetup{
    colorlinks = true,
    urlcolor   = blue,
    citecolor  = black,
}

\newcommand{\RomanNumeralCaps}[1]

\title{Linear stability of cylindrical, multicomponent vesicles}

\author{Anirudh Venkatesh*\aff{1}, Aman Bhargava*\aff{1}, \and Vivek Narsimhan\aff{1}\corresp{\email{vnarsim@purdue.edu}}}

\affiliation{*Equal Contribution\aff{1}Davidson School of Chemical Engineering, Purdue University, West Lafayette, Indiana 47907, USA 
}

\begin{document}
\maketitle

\begin{abstract}
Vesicles are important surrogate structures made up of multiple phospholipids and cholesterol distributed in the form of a lipid bilayer. Tubular vesicles can undergo pearling – i.e., formation of beads on the liquid thread akin to the Rayleigh-Plateau instability. Previous studies have inspected the effects of surface tension on the pearling instabilities of single-component vesicles. In this study, we perform a linear stability analysis on a multicomponent cylindrical vesicle. We solve the Stokes equations along with the Cahn-Hilliard equations to develop the linearized dynamic equations governing the vesicle shape and surface concentration fields. This helps us show that multicomponent vesicles can undergo pearling, buckling, and wrinkling even in the absence of surface tension, which is a significantly different result from studies on single-component vesicles. This behaviour arises due to the competition between the free energies of phase separation, line tension, and bending for this multi-phospholipid system. We determine the conditions under which axisymmetric and non-axisymmetric modes are dominant, and supplement our results with an energy analysis that shows the sources for these instabilities.  We further show that these trends qualitatively match recent experiments \citep{Yanagisawa2010}. 
\end{abstract}

\begin{keywords}
Authors should not enter keywords on the manuscript, as these must be chosen by the author during the online submission process and will then be added during the typesetting process (see \href{https://www.cambridge.org/core/journals/journal-of-fluid-mechanics/information/list-of-keywords}{Keyword PDF} for the full list).  Other classifications will be added at the same time.
\end{keywords}

{\bf MSC Codes}  {\it(Optional)} Please enter your MSC Codes here

\section{Introduction}
\label{sec:introduction}
Vesicles are miniature sacs of fluids surrounded by a thin lipid bilayer, which are often studied to understand the biophysics of cell membranes \citep{LitschelSchwille,Seifert_Lipowsky_1995}. The lipid bilayer demonstrates elasticity that resists changes in dilatation and bending, and these properties makes vesicle dynamics different from conventional fluid droplets \citep{Helfrich_1973,Seifert_1997}.

Vesicles that contain a single lipid species are known as single-component vesicles.  Deflated vesicles of this form demonstrate a wide range of behaviours such as tank treading, tumbling, and trembling under shear flow \citep{Deschamps2009, Vlahovska2007,Abreu_Levant_Steinberg_Seifert_2014}, and stretching instabilities under extensional flow \citep{Narsimhan_Spann_Shaqfeh_2015,boedec_jaeger_leonetti_2014,NarsimhanThesis}. When a tubular vesicle is subject to an external force or perturbation, it may undergo a Rayleigh-Plateau like instability known as `pearling' under tension \citep{BarZiv1994,BarZiv1998, Powers2010,Goldstein1996,boedec_jaeger_leonetti_2014}, and buckling/wrinkling instabilities under compression \citep{Narsimhan_Spann_Shaqfeh_2015} . The pearling phenomenon has been observed for liquid drops \citep{Tomotika1935} and jets \citep{SuryoPOF2006}.  Recently, linear stability analyses have been performed on single-component, tubular vesicles to quantify the onset of pearling, buckling, and wrinkling modes, including the effects of membrane’s bending rigidity, surface viscosity, and applied tension \citep{Narsimhan_Spann_Shaqfeh_2015}. 

In most biological, pharmaceutical, and industrial applications, lipid bilayers contain multiple phospholipids and cholesterol mixtures. These mixtures form phase-separated domains – i.e., lipid rafts -- that are vitally important in signal transduction and protein transport across the cell membrane in biology \citep{Simons_Nature}. This behaviour arises due to the repulsive interactions between saturated and unsaturated lipids on the interface, leading to a liquid-ordered (cholesterol rich) phase and a liquid-disordered (cholesterol poor) phase on the interface \citep{Veatch_Keller,Shimshick_McConnell_1973,Elson_Fried_Dolbow_Genin_2010}. Under these conditions, phase separation on the vesicle surface causes inhomogeneities in material properties like the bending stiffness \citep{Claessens_PRE2007}. These inhomogeneous properties make for interesting physics under flow, and is important in understanding a multitude of physical processes \citep{Baumgart2003,BarthesBiesel2016,gera_salac_spagnolie_2022}.  For example, recent experiments have shown that phase-separated vesicles can give rise to pearling and buckling instabilities \citep{Yanagisawa2010}.   

In this paper, we perform a linear stability analysis of a cylindrical thread with multiple lipids on it, and determine the conditions under which it is unstable under tension or compression.  We will discuss how these results differ from the classical results for a single-component vesicular thread, and perform a qualitative comparison with recent experimental results on multicomponent threads.
 Section \ref{sec:mathematical_formulation} lays out the mathematical formulation of the problem and outlines the characteristic time scales and dimensionless quantities governing the system. This is followed by the linear stability analysis and final reduced equations in section \ref{sec:linstabilityanalysis}. We refresh the memory of the reader by providing results for single-component vesicles in section \ref{sec:single_component}. In section \ref{sec:multicomponent_analysis}, we first provide a general set of observations pertaining to multicomponent vesicles. We then describe the conditions under which one observes axisymmetric versus non-axisymmetric instabilities, and quantify growth rates and dominant wavenumbers. Interestingly, we find that under certain situations, one can observe multimodal instabilties since the growth rates for the axisymmetric and non-axisymmetric modes are comparable.  We provide qualitative comparisons to previous experimental studies \citep{Yanagisawa2010} and quantify the energetic contributions to the stability behaviour. We provide conclusions in section \ref{sec:conclusions}.

\vspace{-2em}
\section{Mathematical Formulation}\label{sec:mathematical_formulation}
\begin{figure}
\centering
\includegraphics[width=\textwidth]{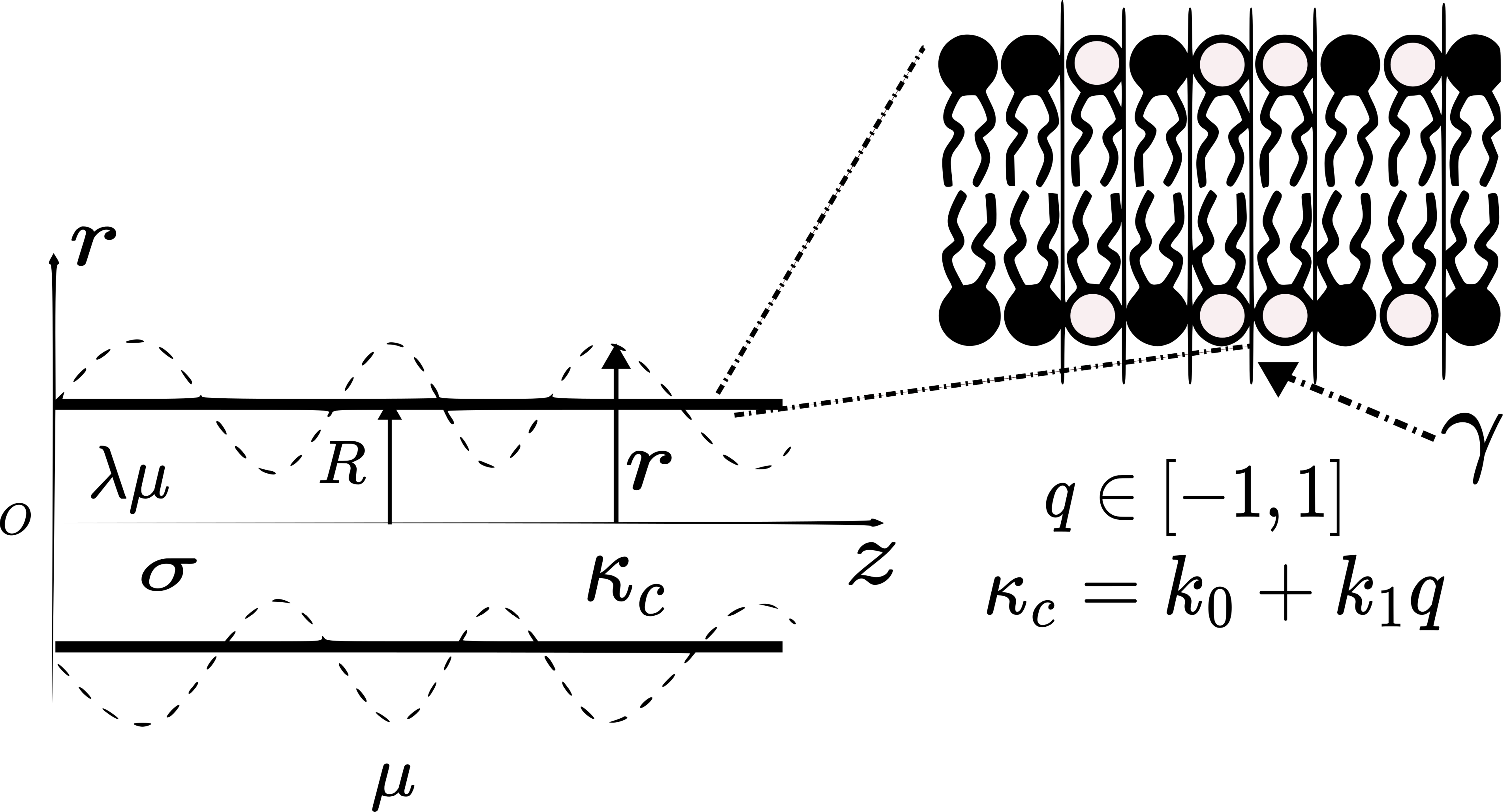}
\caption{Problem setup.  We examine the stability of a cylindrical vesicle with Newtonian fluid inside and outside with viscosities $\lambda \mu$ and $\mu$ respectively.  The membrane has multiple lipids and is characterized by an order parameter $q$ representing different phase-separated domains, a bending modulus $\kappa_c$ depending on $q$, a line tension parameter $\gamma$, and surface tension $\sigma$.}
\label{fig:schematic}
\end{figure}

Figure \ref{fig:schematic} shows an initially cylindrical lipid membrane with Newtonian fluids inside and outside with viscosities $\lambda \mu$ and $\mu$, respectively.  The membrane contains multiple phospholipids that are initially well-mixed, but can potentially phase separate into liquid-ordered ($L_o$) and liquid-disordered domains ($L_d$).  The membrane is incompressible and characterized by an isotropic surface tension $\sigma_0$, a spatially-varying bending modulus $\kappa_c$, and a line tension between the domains (characterized by parameter $\gamma$ described later this section).  We will perform a linear stability analysis by perturbing the membrane shape and lipid concentration, and determine how the shape and phase behaviour evolve over time.  

\subsection{Membrane energy}\label{sec:Membrane_energy}
The energy of the lipid membrane is governed by three factors: bending, phase energy, and surface tension.  The bending energy is given by the classic Canham-Helfrich model  \citep{Helfrich_1973}: 

\begin{equation}\label{eq:bending_Helfrich}
    W_{bend} = \int{\frac{1}{2}\kappa_{c} H^{2}dS}
\end{equation}
In the above equation, $H = \frac{1}{2} \nabla_s \cdot \boldsymbol{n}$ is the mean curvature of the membrane, where $\boldsymbol{n}$ is the outward-pointing normal vector and $\nabla_s = (\boldsymbol{I – nn}) \cdot \nabla$ is the surface gradient operator.  The bending modulus $\kappa_{c}$  depends on the lipid distribution on the membrane.  We represent it as $\kappa_c= \left(\frac{\kappa_{lo}+\kappa_{ld}}{2}\right) + \left(\frac{\kappa_{lo}-\kappa_{ld}}{2}\right)q$, where $\kappa_{lo}$ and $\kappa_{ld}$ are bending moduli of the $L_o$ and $L_d$ phases, and $q$ is an order parameter that represents the phase behaviour of the system ($q = -1$ corresponds to pure $L_d$ phase, while $q = +1$ corresponds to pure $L_o$ phase).  Going forward, we will denote $k_0 = \left(\frac{\kappa_{lo}+\kappa_{ld}}{2}\right)$ as the average bending rigidity and $k_1 =\left(\frac{\kappa_{lo}-\kappa_{ld}}{2}\right)$ as half the bending difference.  Thus, $\kappa_c = k_0 + k_1 q$.  

The order parameter is determined by the thermodynamics of mixing between the membrane’s phospholipids. There are many thermodynamic models available in the literature depending on the specific type of lipids involved and the level of accuracy required \citep{Almeida_2009}. However, the simplest model that qualitatively captures the physics of phase separation is the Landau-Ginzberg equation \citep{safran2018statistical}. Physically,  when one marches along the coordinate that represents a tie line in a phase diagram, the free energy will have two local minima with a barrier in-between if phase separation occurs.  The simplest shape that represents this behaviour is a quartic polynomial, and hence one can write the free energy as

\begin{equation}
W_{phase} = \int{\left(\frac{a}{2}|q|^{2}+\frac{b}{4}|q|^{4} + \frac{\gamma^{2}}{2}|\nabla_{s}q|^{2}\right)dS}
\end{equation}
where $q$ is the order parameter (i.e., coordinate along the tie line for the two phases).  The first two terms in the equation represent a quartic free energy with two minima (i.e., two phases) when $a < 0$, and one minima (i.e., one phase) when $a > 0$.  The last term is the free energy penalty for creating phases that is related to line tension $(\xi^{line})$ and the interface width $(\varepsilon^{width})$:
\begin{equation}\label{eq:linetension_equation}
    \xi^{line} = \frac{2\sqrt{2}}{3b}a^{3/2}\gamma
\end{equation}

\begin{equation}\label{eq:width_equation}
    \varepsilon^{width} = \sqrt{\frac{2\gamma^{2}}{a}}
\end{equation}

The Landau-Ginzberg equation has been used to qualitatively model bilayer membranes \citep{GeraSalac2017}.  Specifically, the symmetric form of Landau-Ginzberg equation listed above gives reasonable estimates for $L_o$/$L_d$ phase-coexistence for the case of 1:1:1 ratio of DOPC:DPPC:cholesterol membranes -- see appendix of \cite{CamleyBrown2014} for the estimated dependence of $a,b$, and $\gamma$ for a specific experimental system ($R\sim O(nm)$). 

The last contribution to the free energy arises from surface tension.  
\begin{equation}
    W_{\sigma} = \int \sigma dS
\end{equation}
Since the number of lipids per unit area is conserved, the membrane surface is incompressible. Thus, $\sigma$ is a Lagrange multiplier used to ensure this constraint.  The surface tension is determined up to an isotropic component $\sigma_0$, which is specified beforehand.  When $\sigma_0 > 0$, the membrane is initially under tension, while when $\sigma_0 < 0$, the membrane is initially under compression. 

\subsection{Dynamical Equations}\label{sec:dyn_eq_full}
We solve the fluid flow inside and outside the membrane in the limit of vanishing Reynolds number.  The Stokes equations are: 
\begin{subequations}
    \begin{equation}
        \mu^{in}\nabla^{2}\boldsymbol{u}^{in} = \nabla p^{in} \quad ; \quad \nabla \cdot \boldsymbol{u}^{in} = 0 \\
    \end{equation}
    \begin{equation}
        \mu^{out}\nabla^{2}\boldsymbol{u}^{out} = \nabla p^{out} \quad ; \quad \nabla \cdot \boldsymbol{u}^{out} = 0
    \end{equation}
\end{subequations}
where $(\boldsymbol{u}, p)$ are the velocity and pressure fields, and $\mu^{in/out}$ are the viscosities inside and outside the vesicle ($\mu^{in} = \lambda \mu$, $\mu^{out} = \mu$).  These equations satisfy continuous velocity across the interface:      
\begin{equation} 
[[\boldsymbol{u}]] = 0 \qquad \boldsymbol{x} \in S 
\end{equation} 
where $[[..]]$ represents the jump across the interface (outer minus inner).  The membrane is surface incompressible: 

\begin{equation}\label{eq:Surface incompressibility} 
\nabla_s \cdot \boldsymbol{u} = 0 \qquad \boldsymbol{x} \in S 
\end{equation} 
where $\nabla_s = (\boldsymbol{I – nn} \cdot \nabla)$ is the surface gradient operator.  Lastly, the hydrodynamic tractions on the interface are balanced by the membrane tractions.    

\begin{equation} \label{eqn:traction_balance}
    \left[\left[\boldsymbol{n}\cdot \boldsymbol{\tau} \right]\right] = \frac{\delta W}{\delta \boldsymbol{x}} \qquad \boldsymbol{x} \in S 
\end{equation} 

In the above equation, $\boldsymbol{\tau}^{in/out} = -p^{in/out}\boldsymbol{I} + \mu^{in/out} \left( \nabla \boldsymbol{u}^{in/out} + \left( \nabla \boldsymbol{u}^{in/out} \right)^T \right)$ is the viscous stress tensor, while the right side is the first variation of the membrane energy with respect to position. This term can be broken into different contributions $\frac{\delta W}{\delta \boldsymbol{x}} = \boldsymbol{f}^{phase} + \boldsymbol{f}^{bend} + \boldsymbol{f}^{\sigma}$, with expressions for each of them listed below:
\begin{subequations}
\begin{equation}
    \boldsymbol{f}^{phase} = \frac{\delta W_{phase}}{\delta \boldsymbol{x}} = -\gamma^2 (\nabla_s^2 q) \nabla_s q - \nabla_s g + 2H \left( \frac{1}{2} \gamma^2 | \nabla_s q |^2 + g \right) \boldsymbol{n}
\end{equation}
\begin{equation} 
    \boldsymbol{f}^{bend} = \frac{\delta W_{bend}}{\delta \boldsymbol{x}} = -\boldsymbol{n} \nabla_{s}^{2}\left(2H \kappa_{c}\right) + \kappa_{c}\left(4HK - 4H^3\right)\boldsymbol{n} - 2H^2 \nabla_s \kappa_c 
\end{equation} 
\begin{equation} 
 \boldsymbol{f}^{\sigma} = \frac{\delta W_{\sigma}}{\delta \boldsymbol{x}} = 2H\sigma \boldsymbol{n} - \mathbf{\nabla}_{s}\sigma 
\end{equation} 
\end{subequations} 
where $g = \frac{a}{2} q^2 + \frac{b}{4}q^4$ is the quartic free energy.  The reader is directed to the following publications for details on how these equations are derived \citep{GeraThesis,Napoli_2010}.  In the above equations, $K = \text{det}(\boldsymbol{L}) = C_{1}C_{2}$ and $H = \frac{1}{2} \text{tr}(\boldsymbol{L}) = \frac{C_{1}+C_{2}}{2}$ are the Gaussian and mean curvatures of the interface respectively, where $\boldsymbol{L} = \nabla_{s}\boldsymbol{n}$ is the surface curvature tensor and $C_{1},C_{2}$ denote the principal curvatures of the interface. The surface tension $\sigma$ is a Lagrange multiplier (up to a specified isotropic constant), which one determines from the surface incompressibility constraint Eq (\ref{eq:Surface incompressibility}) listed above. 

Along with the above flow equations, we also solve a convection-diffusion equation on the vesicle interface for the order parameter $q$.   This equation takes the form of a Cahn-Hilliard equation, the details of which can be found in \cite{GeraThesis}. 

\begin{equation}\label{eq:Cahn_Hilliard_dimensional} 
    \frac{\partial q}{\partial t} + \boldsymbol{u}\cdot \nabla_{s}q = \frac{\nu}{\zeta_{0}} \nabla_{s}^{2}(\zeta) \qquad \boldsymbol{x} \in S 
\end{equation} 

In the above equation, $\nu$ is the characteristic mobility of the phospholipids and $\zeta$ is the surface chemical potential with units of energy per unit area. This chemical potential is the first variation of the membrane energy with respect to the order parameter, while $\zeta_0$ is a reference value provided in \citep{GeraThesis}. 

\begin{equation} \label{eq:chemical_pot}
    \zeta = \frac{\delta W}{\delta q} = aq+bq^{3} - \gamma^{2} \nabla_{s}^{2}q + \frac{k_{1}}{2}(2H)^{2}
\end{equation}

Lastly, the interface satisfies a kinematic boundary condition.  If the vesicle’s shape is characterized by the level set $r = a(z, \phi, t)$, this condition is: 

\begin{equation} 
\frac{ D }{D t} \left(r – a(z,\phi, t) \right)= 0;   \qquad \frac{D}{Dt} = \frac{\partial}{\partial t} + \boldsymbol{u} \cdot \nabla 
\end{equation} 

\subsection{Physical parameters and dimensionless numbers}\label{sec:dimensionless_numbers}
Unless otherwise noted, all remaining quantities in the manuscript will be in dimensionless form.  We nondimensionalize all lengths by cylinder radius $R$, all times by the bending time scale $t_{b} = \mu R^{3}/k_{0}$, and all velocities by $U_b = R/t_{b} = k_{0}/(\mu R^{2})$. All pressures and stresses are scaled by $\mu U_b/R = k_{0}/R^{3}$, and the surface tension is scaled by $k_{0}/R^{2}$. Energies are scaled by $k_0$, and chemical potential is scaled by $k_0/R^2$. 
Table \ref{tbl:Physical_parameter_range} lists the set of physical parameters for this problem and their typical experimental values, while Table \ref{tbl:Dimensionless_parameter_range} lists the dimensionless numbers for this problem.  These dimensionless groups are related to the effects of line tension between the phospholipids, the relative magnitudes of bending stiffness of phospholipids, and size of the vesicle -- depicting an interplay between bending, coarsening, and flow. The most important ones in particular are the viscosity ratio $\lambda$ between the inner and outer fluid, the dimensionless surface tension $\Gamma = \sigma_o R^2/k_0$, the dimensionless bending stiffness difference between the two phases $\beta = k_1/k_0 = (\kappa_{lo} - \kappa_{ld})/(\kappa_{lo}+\kappa_{ld})$, the Cahn number $Cn = \gamma/(R\sqrt{\zeta_0})$ (i.e., ratio of line tension energy to the energy scale of phase separation), the surface Peclet number $Pe = k_0/(\nu \mu R)$ (i.e., {ratio of coarsening time to bending time from diffusion)}, and the line tension parameter $\alpha = k_0/\gamma^2$ (ratio between bending and line tension energies).  Note:  for $\zeta_0 = |a|$ as is the case for most studies, the Cahn number has the alternative interpretation as the ratio of interface width to vesicle radius: 
 $Cn = \varepsilon^{width}/(\sqrt{2} R)$.  See Appendix \ref{app:dimensionlessnumbers} for details.

\begin{table}Physical governing parameters
\centering
  \begin{tabular}{p{0.13\textwidth}p{0.35\textwidth}p{0.29\textwidth}p{0.17\textwidth}}
  \hline
    Variable & Name & Order of Magnitude & Reference\\ 
    $L$ & Length of cylindrical vesicle & $\sim 30 \mu m$ & \cite{Kanstler_Steinberg_2008}\\
    $R$ & Radius of cylindrical vesicle & $\sim 5 \mu m$ & \cite{Kanstler_Steinberg_2008}\\
    $k_{0}$ & Bending stiffness sum between Phospholipid $1$ and $2$ & $O(10^{-19}-10^{-18})J$ & \cite{Amazon2013}\\
    $k_{1}$ & Bending stiffness difference between Phospholipid $1$ and $2$ & $O(10^{-19})J$ & \cite{Amazon2013}\\
    $\nu$ & Mobility of phospholipids & $O(10^{-11})m^{2}/s$ & \cite{Negishi2008}\\
    $\gamma$ & Line tension parameter& $O(10^{-9}) J^{1/2}$& \cite{Luo_Maibaum_2020}\\
    \hline
\end{tabular}%
\vspace{-1em}
  \caption{Physical parameter ranges and orders of magnitude} 
\label{tbl:Physical_parameter_range}
\end{table}

\begin{table}
\centering
  \begin{tabular}{p{0.33\textwidth}p{0.43\textwidth}p{0.2\textwidth}}
    Variable & Name & Order of Magnitude\\ 
    \hline
    $L/R$ & Length to radius ratio & $\sim 5$ \\
    $\Tilde{a} = a/\zeta_{0}$ & Dimensionless double well potential term & $-1$ \\
    $\Tilde{b} = b/\zeta_{0}$ &  Dimensionless double well potential term & $O(1)$ \\
    $\beta = k_{1}/k_{0}$ & Ratio of bending stiffnesses & $O(0.1-1)$  \\
    $Cn = \gamma/(R\sqrt{\zeta_{0}}) = \varepsilon^{width}/(\sqrt{2}R)$ & Cahn number & $O(0.1-1)$ \\
    $\alpha = k_{0}/\gamma^{2}$ & Ratio of bending stiffness to line tension & $O(1)$ \\
    $\lambda $ & Viscosity Ratio & $O(1-10)$ \\
    $Pe = k_{0}/(\nu\mu R)$ & Peclet number (coarsening timescale/bending timescale) & $O(1)$\\
    $\Gamma = \sigma_{0}R^{2}/k_{0}$ & Dimensionless isotropic membrane tension & $O(1-10)$\\
    \hline
\end{tabular}%
\vspace{-1em}
  \caption{Dimensionless parameter ranges and orders of magnitude} 
\label{tbl:Dimensionless_parameter_range}
\end{table}

\vspace{-2em}
\section{Linear stability analysis}\label{sec:linstabilityanalysis}
\subsection{Derivation}
We consider a vesicle that has its base state equal to that of a cylinder at rest (i.e., $r_0 = 1, \boldsymbol{u}^{in}_{0} = \boldsymbol{u}^{out}_0 =0$).  The membrane is uniformly mixed as one phase with an equal amount of stiff and soft lipids (i.e., $q_0 = 0$).  The membrane tension is uniform with a non-dimensional value  $\Gamma = \sigma_{0}R^2/k_0$.  The base pressure inside and outside the cylinder is given by the Young-Laplace law with bending rigidity, which corresponds to $p^{out}_0 = 0, p^{in}_0 = \Gamma – \frac{1}{2}$.  

We perform a linear stability analysis on this base state.  We perturb all geometric and physical quantities an infinitesimal amount $\epsilon \ll 1$ as shown below:
\begin{subequations}
    \begin{align}    
        r &= 1 + \epsilon 
        r_{kn}\exp(ikz+in\phi)\\ 
        \sigma &= \Gamma + \epsilon 
        \sigma_{kn}\exp(ikz+in\phi)\\
        \textbf{u}^{in/out} &=\epsilon\textbf{u}_{kn}^{in/out} \exp(ikz+in\phi)\\
        q &= \epsilon q_{kn}\exp(ikz+in\phi)\\
        p^{in} &= \Gamma - \frac{1}{2} + \epsilon p_{kn}^{in}\exp(ikz+in\phi)\\
        p^{out} &= \epsilon p_{kn}^{out}\exp(ikz+in\phi)
    \end{align}  
\end{subequations}

We then solve the Stokes equations and Cahn-Hilliard equations, linearized to $O(\epsilon)$, and determine how the radius $r$ and concentration field $q$ evolve over time.  The thread is considered unstable if a perturbation causes the radius and concentration to grow over time.  Due to the geometric nature of the problem, all perturbations are decomposed into Fourier modes, where $k$ and $n$ represent axial and azimuthal wavenumbers. 

The first step we perform is to linearize the Cahn-Hilliard equation (Eqs. \ref{eq:Cahn_Hilliard_dimensional} and \ref{eq:chemical_pot}).  Doing so yields a differential equation for the order parameter $q_{kn}$:
\begin{equation}\label{eq:dynamic_composition_IntPe}
    F_{kn} \Dot{q}_{kn} = M_{kn} r_{kn} + V_{kn} q_{kn}
\end{equation}
In the above equation, the right hand side is equal to the linearized chemical potential $\delta W/\delta q$, while the left hand side is a dynamical factor.  The coefficients are given by:

\begin{subequations}
    \begin{equation} \label{eq:Fkn}
        F_{kn} = -\frac{Pe}{Cn^2 \alpha} \frac{1}{k^2 + n^2}
    \end{equation}

    \begin{equation} \label{eq:Mkn}
        M_{kn} = \beta \left( k^2 + n^2 - 1 \right)
    \end{equation}
    
    \begin{equation} \label{eq:Vkn}
        V_{kn} = \frac{1}{Cn^2 \alpha} \left[ \Tilde{a} + Cn^2 \left(k^2 + n^2 \right) \right]
    \end{equation}

\end{subequations}

To obtain the differential equation for the vesicle shape $r_{kn}$, we follow a procedure similar the previous publications for single-component vesicles (see \cite{NarsimhanThesis, Narsimhan_Spann_Shaqfeh_2015}).  First, we solve the Stokes equations inside and outside the vesicle.  We use the cylindrical harmonics solution given in \cite{Happel_Brenner}:

\begin{subequations}\label{eq:cylindrical_harmonics}
    \begin{equation}
        \mathbf{u}_{kn}\exp({ikz+in\phi}) = \nabla \psi + \nabla \times (\Omega \hat{\mathbf{z}}) + r\frac{\partial}{\partial r}(\nabla \Pi) + \mathbf{\hat{z}}\frac{\partial \Pi}{\partial z}
    \end{equation}
    \begin{equation}
        p_{kn}\exp({ikz+in\phi}) = -2\Tilde{\eta}\frac{\partial^{2}\Pi}{\partial z^{2}}
    \end{equation}
\end{subequations}
where $\Tilde{\eta}$ is the non-dimensional viscosity ($\Tilde{\eta} = 1$ outside the vesicle and $\Tilde{\eta} = \lambda$ inside) and $\psi,\Omega,$ and $\Pi$ are scalar harmonic functions:

\begin{equation}
    {\{\psi,\Omega,\Pi\}} = {\{A_{kn},iB_{kn},C_{kn}\}}G_{n}(kr)\exp(ikz+in\phi)
\end{equation}

In the above equation, the functions $G_{n}(kr)$ are modified Bessel functions, equal to $I_{n}(kr)$ inside the vesicle and $(-1)^n K_{n}(kr)$ outside the vesicle.  Writing the velocity and pressure fields in this form yields seven unknowns for each Fourier mode, which we solve through appropriate boundary conditions.  The unknowns are the coefficients $\{ A_{kn}^{out}, B_{kn}^{out}, C_{kn}^{out} \}$ outside the vesicle, the coefficients $\{ A_{kn}^{in}, B_{kn}^{in}, C_{kn}^{in} \}$ inside the vesicle, and the non-isotropic surface tension $\sigma_{kn}$ that arises from membrane incompressibility.

Below is the structure of the linear equations we solve.  The structure is given by $\boldsymbol{W} \cdot \boldsymbol{y}=\boldsymbol{b}$, where $\boldsymbol{W}$ is a matrix, $\boldsymbol{y} =\{ A_{kn}^{out}, B_{kn}^{out}, C_{kn}^{out}, A_{kn}^{in}, B_{kn}^{in}, C_{kn}^{in}, \sigma_{kn}^M \}$ is the vector of unknowns where $\sigma_{kn}^M = \sigma_{kn} + \frac{ \beta}{2} q_{kn}$ is a modified surface tension, and $\boldsymbol{b}$ is the right hand side.  We use a modified surface tension for convenience since the linear system below becomes exactly the same as in previous literature for single-component vesicles \citep{Narsimhan_Spann_Shaqfeh_2015}.

\begin{multline} \label{eq:matrix_eq}
    \begin{bmatrix}
        W_{11} & W_{12} & W_{13} & W_{14} & W_{15} & W_{16} & W_{17} \\
        W_{21} & W_{22} & W_{23} & W_{24}& W_{25} & W_{26} & W_{27} \\
        W_{31} & W_{32} & W_{33} & W_{34} & W_{35}& W_{36} & W_{37} \\
        W_{41} & W_{42} & W_{43} & W_{44} & W_{45} & W_{46} & W_{47} \\
        W_{51} & W_{52} & W_{53} & W_{54} & W_{55} & W_{56} & W_{57} \\
        W_{61} & W_{62} & W_{63}& W_{64}& W_{65} & W_{66}& W_{67} \\
        W_{71}& W_{72}& W_{73} & W_{74} & W_{75} &W_{76} & W_{77} \\
    \end{bmatrix} \begin{bmatrix}
        A^{in}_{kn} \\
        B^{in}_{kn} \\
        C^{in}_{kn} \\
        A^{out}_{kn} \\
        B^{out}_{kn} \\
        C^{out}_{kn} \\
        \sigma_{kn}^{M} \\
    \end{bmatrix} 
    = \begin{bmatrix}
        b_1 \\
        b_2 \\
        b_3 \\
        b_4 \\
        b_5 \\
        b_6 \\
        b_7 \\
    \end{bmatrix}
\end{multline}

In the above linear system, each row arises from a boundary condition. The entries are summarized below, where $I_n$ and $K_n$ are evaluated at wavenumber $k$ and $I_n', I_n'', K_n', K_n''$ are single and double derivatives evaluated at $k$.  The entries below are exactly the same as those found in the prior literature.

\begin{itemize}
    \item \textbf{Row 1:  Continuity of velocity ($[[u_z]]=0$ at $r=1$)}
    \begin{equation}
    \begin{split}
&W_{11} = -kI_{n}; \quad  W_{12} = 0; \quad W_{13} = -k^{2} I_{n}' - kI_{n}; \quad W_{14} = (-1)^n kK_{n};\\
& W_{15} = 0; \quad W_{16} = (-1)^{n}k^2 K_{n}'+(-1)^n kK_{n}; \quad  W_{17} = 0; \quad b_1 = 0
\end{split}
    \end{equation}

    \item \textbf{Row 2:  Continuity of velocity ($[[u_{\phi}]]=0$ at $r=1$)}
    \begin{equation}
    \begin{split}
&W_{21} = -n I_{n}; \quad W_{22} = kI_{n}'; \quad W_{23} = -nk I_{n}'+nI_{n}; \quad W_{24} = (-1)^{n} nK_{n}\\
&W_{25} = (-1)^{n+1}k K_{n}'; \quad W_{26} = (-1)^{n} nk K_{n}'-(-1)^{n}nK_{n}; \quad W_{27} = 0; \quad b_2 = 0
\end{split}
    \end{equation}

    \item \textbf{Row 3:  Kinematic boundary condition ($u_r^{in}=\frac{dr}{dt}$ at $r=1$)}
    \begin{equation}
    \begin{split}
&W_{31} = k I_{n}'; \quad W_{32} = -n I_{n}; \quad W_{33} = k^2 I_{n}''; \quad W_{34} = 0\\
&W_{35} = 0; \quad W_{36} = 0; \quad W_{37} = 0; \quad b_3 = \Dot{r}_{kn}
\end{split}
    \end{equation}
       
    \item \textbf{Row 4:  Kinematic boundary condition ($u_r^{out}=\frac{dr}{dt}$ at $r=1$)}
    \begin{equation}
    \begin{split}
&W_{41} = 0; \quad W_{42} = 0; \quad W_{43} = 0; \quad W_{44} = (-1)^{n} k K_{n}'\\
&W_{45} = (-1)^{n+1} n K_{n}; \quad W_{46} = (-1)^{n} k^2 K_{n}''; \quad W_{47} = 0; \quad b_4 = \Dot{r}_{kn}
\end{split}
    \end{equation}

    \item \textbf{Row 5:  Surface incompressibility ($\nabla_s \cdot \boldsymbol{u}^{out} =0$ at $r=1$)}
    \begin{equation}
    \begin{split}
&W_{51} = W_{52} = W_{53} = 0; \quad W_{54} = (-1)^{n} \left( k K_n' - (n^2 + k^2)K_n \right) \\ 
&W_{55} = (-1)^{n} \left(-n K_n + kn K_n' \right) ; \\
&W_{56} = (-1)^{n} \left( k^2 K_n'' - k (n^2 + k^2) K_n' + (n^2 - k^2) K_n \right); \quad W_{57} = 0; \quad b_5 = 0
\end{split}
    \end{equation}

     \item \textbf{Row 6:  Tangential stress balance ($[[\tau_{zr}]] + \frac{\partial \sigma^M}{\partial z} = 0$ at $r=1$)}
    \begin{equation}
    \begin{split}
&W_{61} = -2 \lambda k I_n'; \quad W_{62} = \lambda n I_n ; \quad W_{63} = -\lambda \left( 2k^2 I_n'' + 2 k I_n'\right); \\ 
&W_{64} = (-1)^n 2k K_n';  \quad W_{65} = (-1)^{n+1} n K_n ; \quad W_{66} = (-1)^n \left( 2k^2 K_n'' + 2 k K_n'\right); \\
&W_{67} = 1; \quad b_6 = 0
\end{split}
    \end{equation}

    \item \textbf{Row 7:  Tangential stress balance ($[[\tau_{\phi r}]] + \frac{1}{r}\frac{\partial \sigma^M}{\partial \phi} = 0$ at $r=1$)}
    \begin{equation}
    \begin{split}
&W_{71} = -\lambda \left( 2nk I_n' - 2n I_n\right); \quad W_{72} = -\lambda \left( -n^2 I_n + k I_n' - k^2 I_n''\right);  \\
& W_{73} = -\lambda \left( 2nk^2 I_n'' - 2nk I_n' + 2n I_n \right); \quad W_{74} = (-1)^n \left( 2nk K_n' - 2n K_n\right); \\
&W_{75} = (-1)^{n} \left( -n^2 K_n + k K_n' - k^2 K_n''\right); \quad W_{76} = (-1)^n \left( 2nk^2 K_n'' - 2nk K_n' + 2n K_n \right); \\
&W_{77} = n; \quad b_7 = 0
\end{split}
    \end{equation}

\end{itemize}

After we solve for the unknowns, we apply the last boundary condition -- the normal stress balance -- to obtain the final differential equation for the vesicle shape.  The linearized normal stress boundary condition (Eq \ref{eqn:traction_balance}) is:

\begin{equation}\label{eq:normal_stress_Oeps}
    -[[p_{kn}]] - \sigma_{kn}^M = L_{kn}r_{kn} + M_{kn} q_{kn}
\end{equation}
where the left hand side comes from the pressure and surface tension obtained from the unknowns solved above, and the right hand side comes from the linearized membrane traction $\boldsymbol{f} = \delta W/\delta \boldsymbol{x}$ (minus the modified surface tension contribution $\sigma_{kn}^M$).  The expression for $M_{kn}$ is the same as in Eq. (\ref{eq:Mkn}), while $L_{kn}$ is:

\begin{equation} \label{eq:Lkn}
    L_{kn} = \Gamma \left( n^{2}+k^{2}-1 \right)+\frac{3}{2} + 2k^{2} + \left(n^{2}+k^{2}\right) \left(n^{2}+k^{2}-\frac{5}{2} \right)
\end{equation}

The expression for the left hand side in Eq. (\ref{eq:normal_stress_Oeps}) in terms of the solved coefficients is $-[[p_{kn}]] - \sigma_{kn}^M = 2k^2 \left( \lambda I_n C_{kn}^{in} + (-1)^{n+1} K_n C_{kn}^{out} \right) - \sigma_{kn}^M$.  Since the latter quantities are linear in the rate of interface deformation $\Dot{r}_{kn}$, we can rewrite the above expression (Eq \ref{eq:normal_stress_Oeps}) as:

\begin{equation}\label{eq:eqn_rkn}
    \Lambda_{kn} \Dot{r}_{kn} = L_{kn}r_{kn} + M_{kn} q_{kn}
\end{equation}

This equation (Eq. \ref{eq:eqn_rkn}) along with the linearized Cahn-Hilliard equation (Eq. \ref{eq:dynamic_composition_IntPe}) are the dynamical equations obtained for the linear stability analysis.  In general, there is no analytical solution for the coefficient $\Lambda_{kn}$ -- it must be computed numerically by inverting the system of equations (\ref{eq:matrix_eq}).  However, for the specific case of axisymmetric modes $(n=0)$, analytical expressions are available; details are provided in the appendix - section \ref{appsec:axisymmetric}.  

\subsection{Final structure of equations}\label{sec:final_structure_eqns}
The final form of the dynamical equations are:

\begin{equation} \label{eq:final_eq}
\begin{bmatrix}
    \Lambda_{kn} & 0 \\
    0 & F_{kn} \\         
\end{bmatrix} \cdot \frac{d}{d t}
\begin{bmatrix}
    r_{kn} \\
    q_{kn} \\         
\end{bmatrix} = 
\begin{bmatrix}
    L_{kn} & M_{kn} \\
    M_{kn} & V_{kn} \\         
\end{bmatrix} \cdot
\begin{bmatrix}
    r_{kn} \\
    q_{kn} \\         
\end{bmatrix}
\end{equation}
where entries $\Lambda_{kn}$, $F_{kn}$, $L_{kn}$, $M_{kn}$, and $V_{kn}$ were described in the previous section (see Eqs. (\ref{eq:Fkn})-(\ref{eq:Vkn}), (\ref{eq:Lkn}), and text below (\ref{eq:Lkn})).  A few comments are made here:

\begin{itemize}
    \item The left hand side entries $\Lambda_{kn}$ and $F_{kn}$ are purely dynamical quantities that depend on the hydrodynamics of the surrounding fluid as well as the diffusion characteristics of the lipids.  They are negative definite -- i.e., $\Lambda_{kn}, F_{kn} < 0$, so they do not alter the stability of the system, but play a role in the timescale of the instability as well as mode selection.  $\Lambda_{kn}$ depends on the viscosity ratio $\lambda$, while $F_{kn}$ depends on the quantity $Pe/(\alpha Cn^2)$, which equals the diffusion time divided by the chemical potential relaxation time. 
    
    \item The right hand side entries $L_{kn}, M_{kn}, V_{kn}$ are related to the second variation in the free energy at the base state $r_{kn}, q_{kn} = 0$:

    \begin{equation}
        \begin{bmatrix}
        L_{kn} & M_{kn} \\
        M_{kn} & V_{kn} \\         
        \end{bmatrix}
        \sim
        \begin{bmatrix}
        \frac{\partial^2 W}{\partial r_{kn} \partial r_{kn}} & \frac{\partial^2 W}{\partial r_{kn} \partial q_{kn}} \\
        \frac{\partial^2 W}{\partial r_{kn} \partial q_{kn}} & \frac{\partial^2 W}{\partial q_{kn} \partial q_{kn}} \\      
        \end{bmatrix} 
    \end{equation}
    Thus, the matrices are only related to the elastic and mixing energies of the system, and depend only on quantities related to the bending moduli, surface tension, line tension, and quartic energy potential.  Since these matrices are related to the local curvature of the free energy landscape, the sign of eigenvalues determine the relative stability of the system.  For example, if the energy is concave down, the system is unstable. \\
\end{itemize}

\subsection{Modal analysis}\label{sec:modal_analysis}

We will perform an eigenvalue/eigenvector analysis on the ODEs in  (\ref{eq:final_eq}).  For each set of wavenumbers $(k, n)$, we will write the system of equations in the form $\boldsymbol{\dot{y}} = \boldsymbol{M} \cdot \boldsymbol{y}$, where $\boldsymbol{y} = [r_{kn}, q_{kn}]$, and then obtain the two eigenvalue/eigenvector pairs for the matrix $\boldsymbol{M}$.  The shape is considered to be unstable if there is at least one eigenpair that has a positive eigenvalue and a non-zero component in the $r_{kn}$ direction.  The most dangerous of the two eigenpairs is the one that has the largest eigenvalue.  
 
 We denote the growth rate $s$ for a given wavenumber $(k, n)$ as the largest eigenvalue:
 
 \begin{equation}
     s = \max{ \text{eig}(\boldsymbol{M}})
 \end{equation}
 We will determine the range of wavenumbers that lead to instability by obtaining the set of $(k,n)$ that lead to a positive growth rate.  The most dangerous mode $(k_{max}, n_{max})$ is determined by finding $(k,n)$ that maximize the growth rate.  Unlike the single-component vesicle case where only the axisymmetric $(n = 0)$ modes are unstable under tension, the multicomponent case can have non-axisymmetric modes ($n > 1$) being unstable; thus, we will examine a wide range of values $(n,k)$ in this paper and commment on the type of instabilities formed.
 

\section{Single-component analysis}\label{sec:single_component}

In this section, we review prior literature on single-component vesicles and validate our equations against published results.  

\begin{figure}
     \centering
     \begin{subfigure}{0.32\textwidth}
       \includegraphics[width=\textwidth]{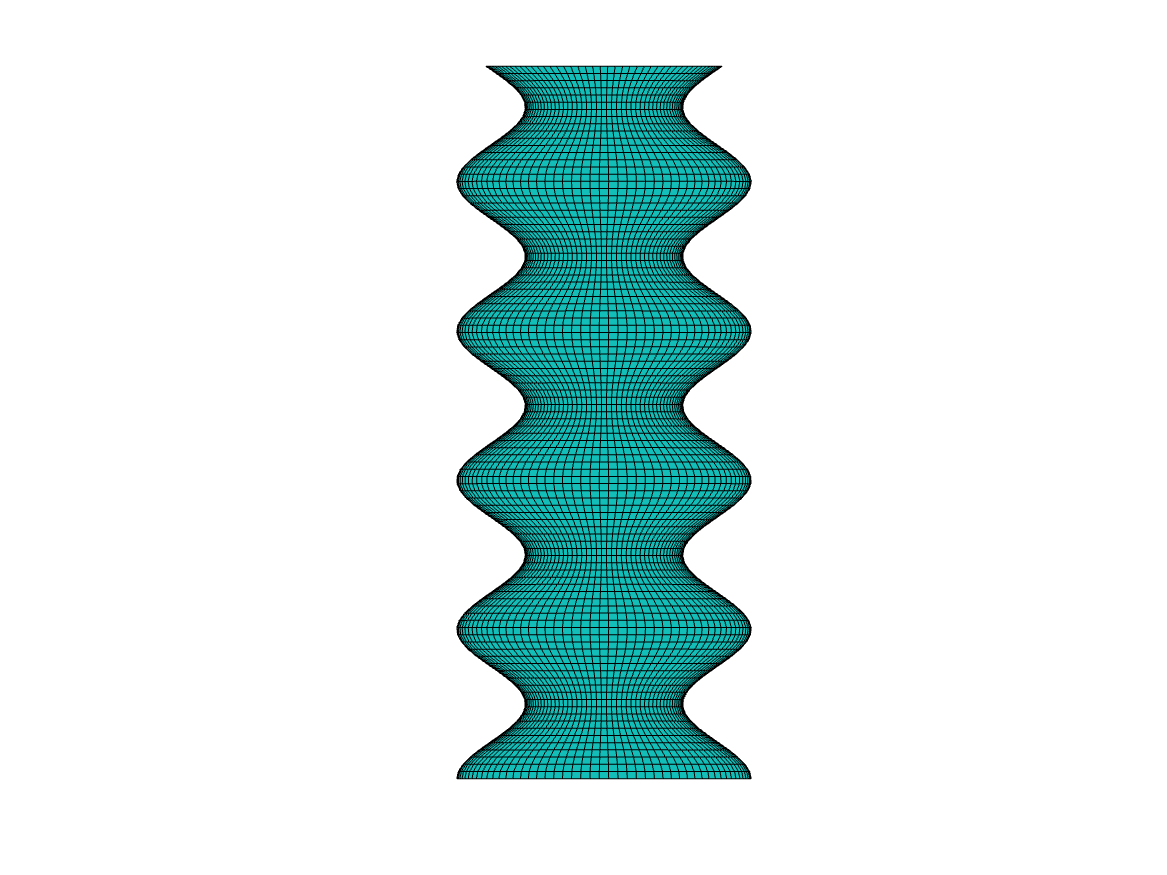}
       \captionsetup{font=normalsize,labelfont={bf,sf}}
       \caption{Pearling}
       \label{fig:Pearling_Visualization_Single}
     \end{subfigure}
     \begin{subfigure}{0.32\textwidth}              
        \includegraphics[width=\textwidth]{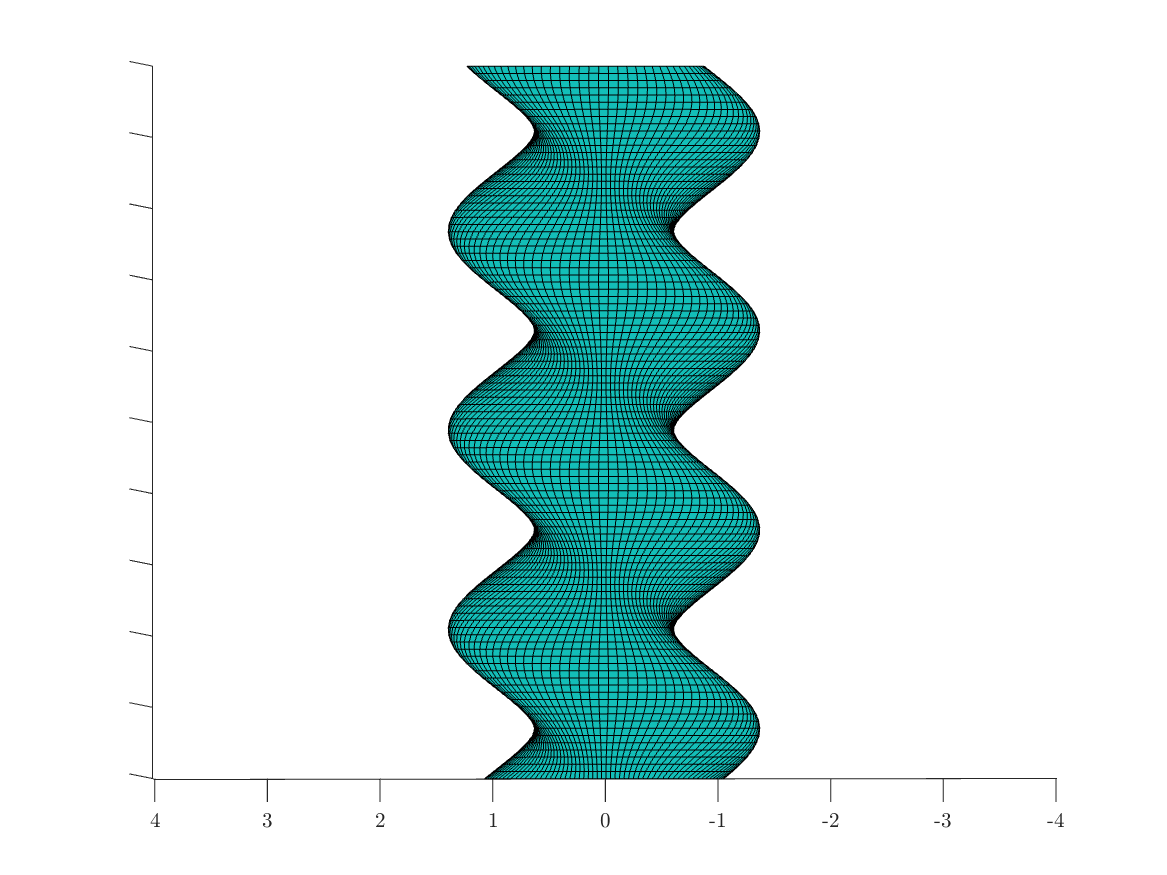}
        \captionsetup{font=normalsize,labelfont={bf,sf}}
        \caption{Buckling}
       \label{fig:Buckling_Visualization_Single}
     \end{subfigure}
    \begin{subfigure}{0.32\textwidth}              
        \includegraphics[width=\textwidth]{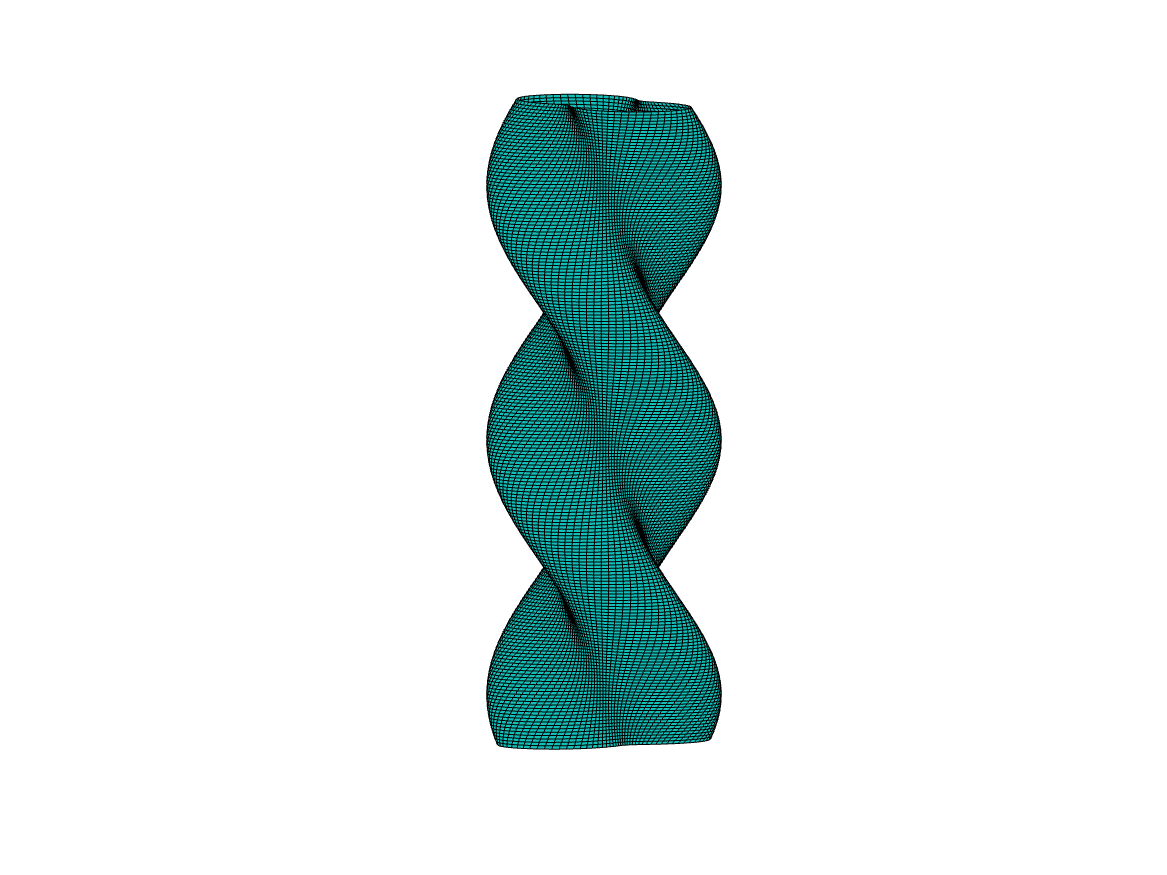}
        \captionsetup{font=normalsize,labelfont={bf,sf}}
        \caption{Wrinkling}
       \label{fig:Wrinkling_Visualization_Single}
     \end{subfigure}
     \caption{Snapshots of $(a)$ pearling $(n=0)$ $(b)$ buckling $(n=1)$ $(c)$ wrinkling $(n=2)$ modes for single-component vesicles.}
     \label{fig:Single_Component_Visualization}
\end{figure}

For single-component lipid threads, the formation of instabilities depends on one control parameter, the non-dimensionalized surface tension $\Gamma = \sigma_0 R^2/k_0$. Figure \ref{fig:Single_Component_Visualization} shows pictures of what the instabilities look like.  For this paper, we will coin $n = 0$ modes pearling, $n = 1$ modes as buckling, and $n > 1$ modes as wrinkling.

\begin{figure}
    \centering
    \begin{subfigure}{0.48\textwidth}        \includegraphics[width=\textwidth]{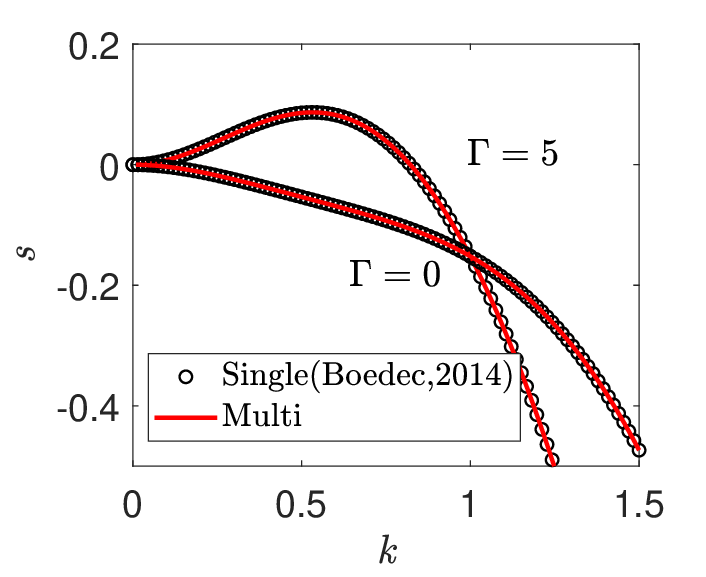}
        \captionsetup{font=normalsize,labelfont={bf,sf}}
        \caption{$n=0$}
        \label{fig:validation_axisymmetric}
    \end{subfigure}
    \begin{subfigure}{0.48\textwidth}
        \includegraphics[width=\textwidth]{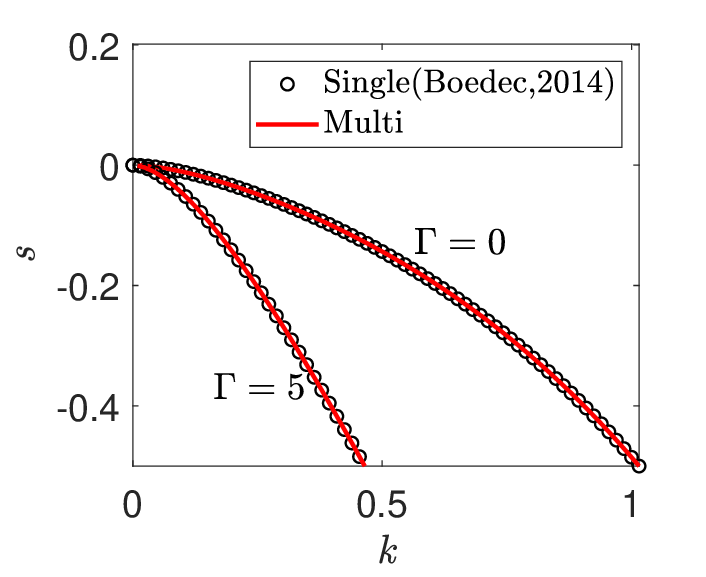}
        \captionsetup{font=normalsize,labelfont={bf,sf}}
        \caption{$n=1$}
        \label{fig:validation_nonaxisymmetric}
    \end{subfigure}
    \caption{Growth rate vs. wavenumber for an equiviscous ($\lambda = 1$), single-component vesicle at $\Gamma=0$ and $\Gamma=5$ for (a) pearling mode ($n=0$), and (b) buckling mode ($n=1$).  Results are validated against published results 
 \citep{boedec_jaeger_leonetti_2014}}
    \label{fig:validation}    
\end{figure}

Figures \ref{fig:validation_axisymmetric} and \ref{fig:validation_nonaxisymmetric} compare the growth rates for the pearling and buckling modes from our theory against published results in the literature for single-component vesicles \citep{boedec_jaeger_leonetti_2014,Narsimhan_Spann_Shaqfeh_2015}.  We obtain single-component results by setting $\beta = 0$, i.e., both phases have same bending rigidities; $Cn = 0$, which corresponds to zero line tension between the phases; and the double-well potential parameter $\tilde{a} = 0$, which ensures that no phase separation occurs. We find an excellent agreement between the growth rates from our analysis with those published previously.  



\begin{figure}
\centering

\includegraphics[width=0.7\textwidth]{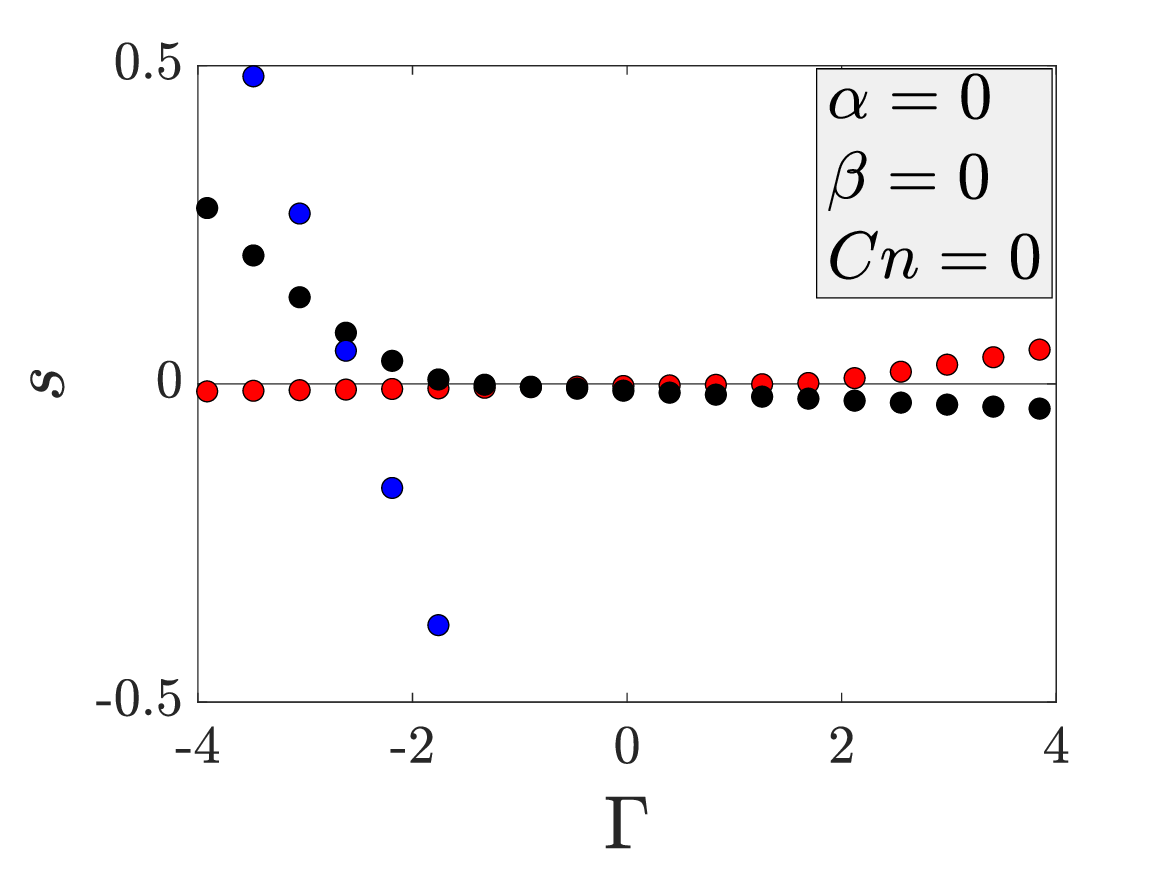}

\caption{Most unstable growth rates with respect to the isotropic membrane tension $\Gamma$ for single-component vesicles. The red circles represent $n=0$ pearling modes, black circles represent $n=1$ buckling modes, and blue circles represent $n=2$ wrinkling modes. In the plot, $\lambda =1$.}
\label{fig:Most_Unstable_wrt_Gamma_Single}
\end{figure}

{Figure \ref{fig:Most_Unstable_wrt_Gamma_Single} presents the most unstable growth rates for the three modes $n=0,1,2$ for different values of the isotropic membrane tension $\Gamma$. If the vesicle is under tension ($\Gamma > 0$), the vesicle is stable to all perturbations for tension values $0 < \Gamma < 3/2$.  When the tension is above a critical value $\Gamma > 3/2$, axisymmetric pearling instabilities ($n = 0$) are unstable (i.e., $s > 0$) and non-axisymmetric modes $n > 0$ are stable.  When the thread is under compression ($\Gamma < 0$), both axisymmetric $n = 0$ and non-axisymmetric modes $n > 0$ modes can become unstable.  The axisymmetric (pearling) mode is unstable for $\Gamma < -(3 + 4\sqrt{2})/2$, the $n = 1$ (buckling) mode is unstable for $\Gamma < -3/2$, and $n > 1$ (wrinkling) modes are unstable for $\Gamma < -(n^2 -3/2)$ \citep{boedec_jaeger_leonetti_2014,Narsimhan_Spann_Shaqfeh_2015}.}

\section{Multicomponent analysis}\label{sec:multicomponent_analysis}

\subsection{General observations and choice of parameter space}
\begin{figure}
     \centering
     \begin{subfigure}{0.33\textwidth}
       \includegraphics[width=\textwidth]{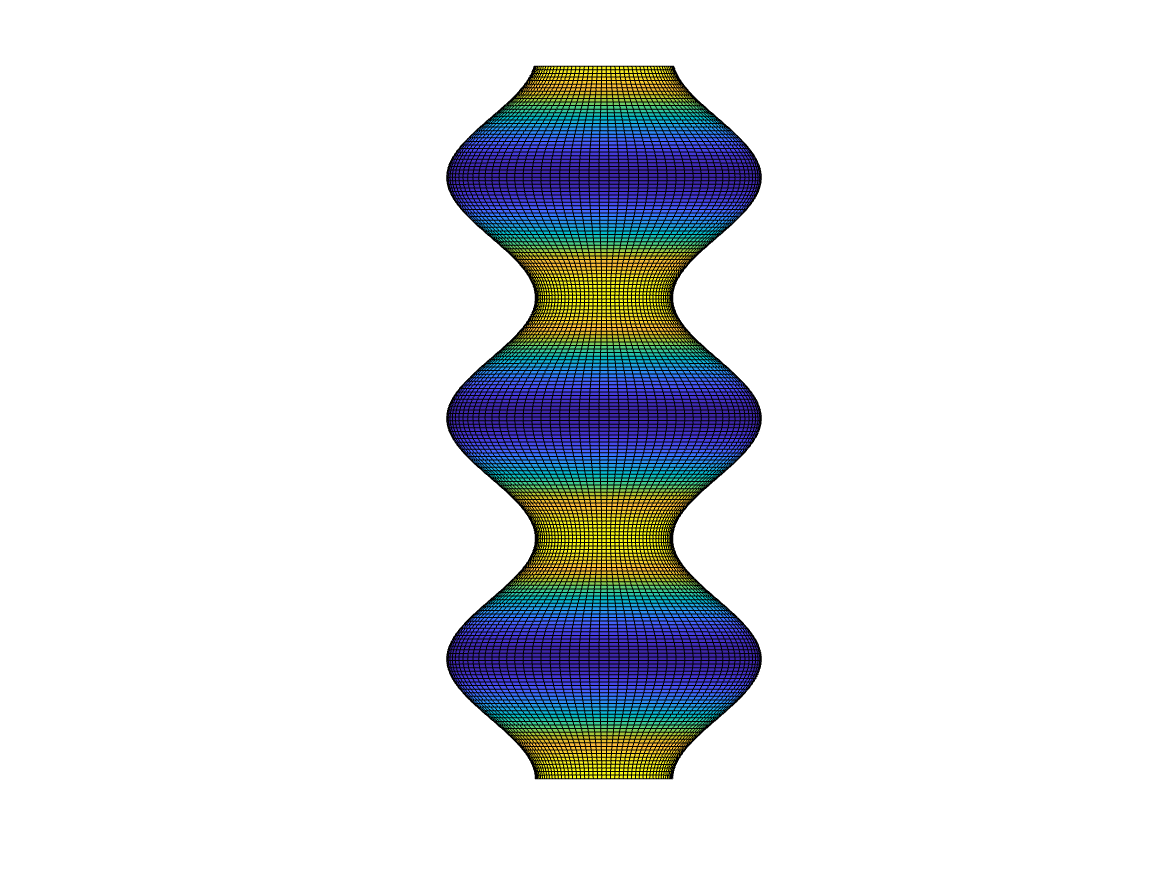}
       \captionsetup{font=normalsize,labelfont={bf,sf}}
       \caption{Pearling $(n=0)$}
       \label{fig:Pearling_Visualization_Multi}
     \end{subfigure}
     \begin{subfigure}{0.33\textwidth}              
        \includegraphics[width=\textwidth]{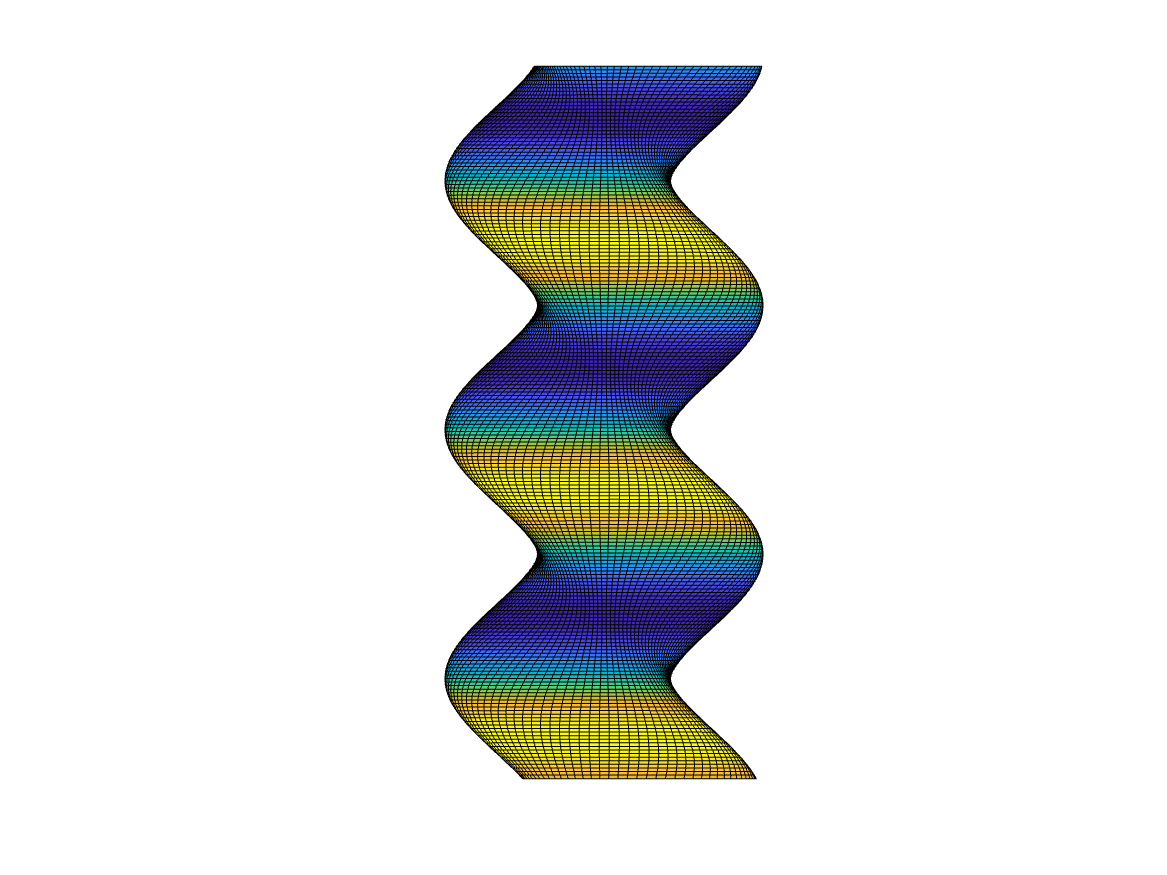}
        \captionsetup{font=normalsize,labelfont={bf,sf}}
        \caption{Buckling $(n=1)$}
       \label{fig:Buckling_Visualization_Multi}
     \end{subfigure}
    \begin{subfigure}{0.33\textwidth}              
        \includegraphics[width=\textwidth]{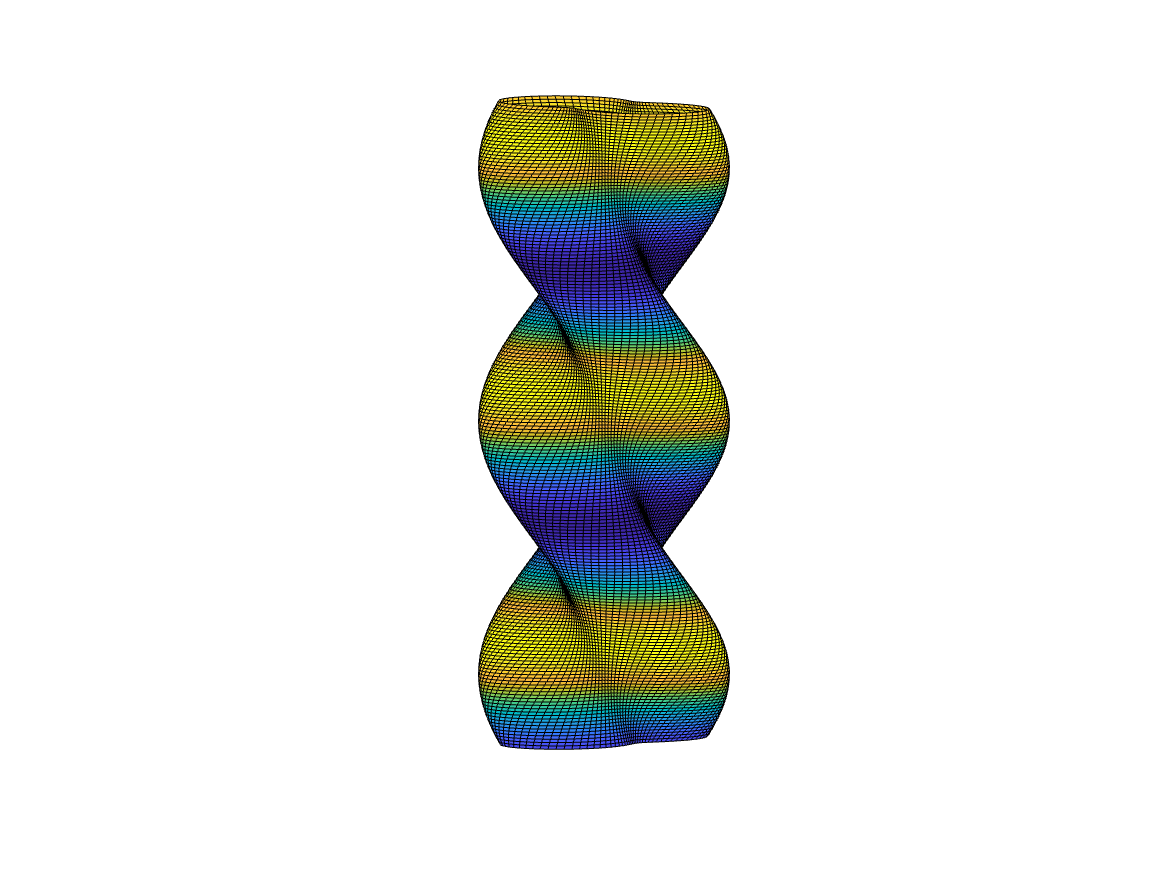}
        \captionsetup{font=normalsize,labelfont={bf,sf}}
        \caption{Wrinkling $(n=2)$}
       \label{fig:Wrinkling_Visualization_Multi}
     \end{subfigure}
\quad
     \begin{subfigure}{0.33\textwidth}              
        \includegraphics[width=\textwidth]{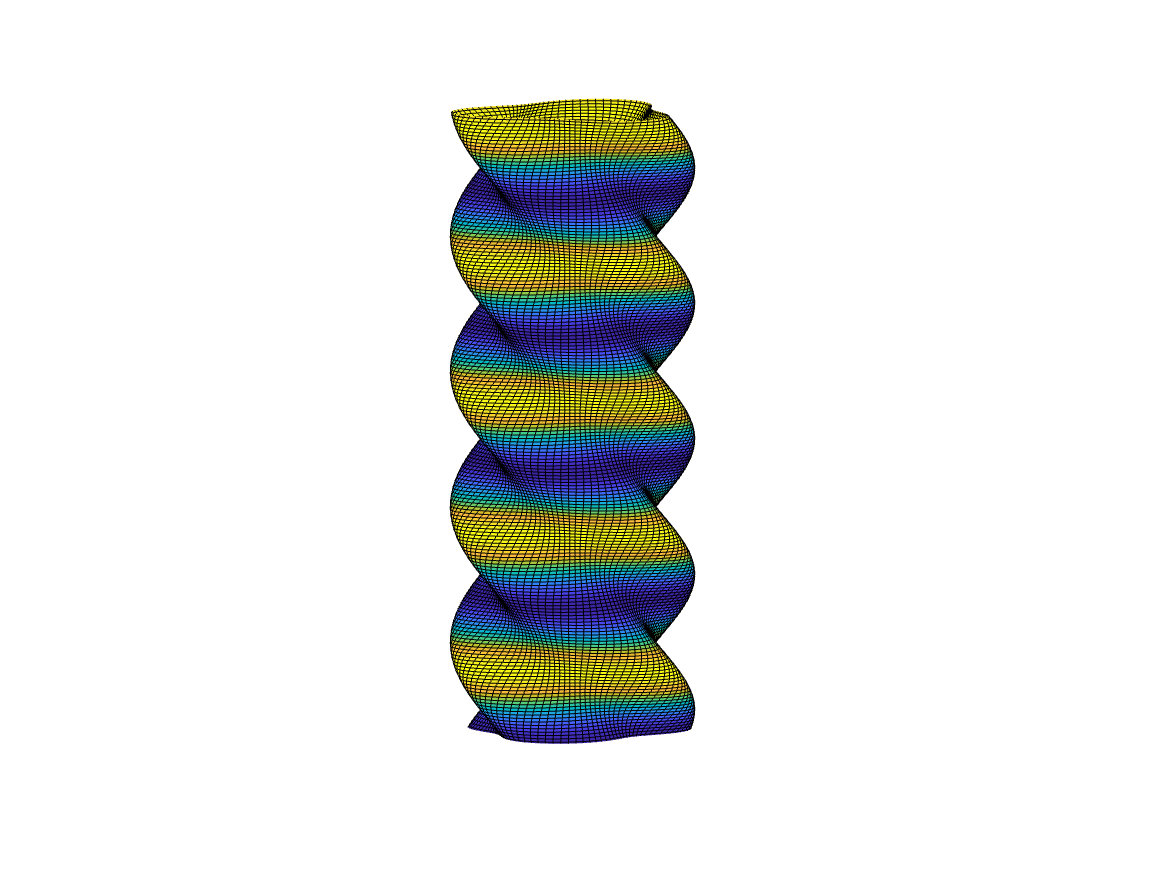}
        \captionsetup{font=normalsize,labelfont={bf,sf}}
        \caption{Wrinkling $(n=3)$}
       \label{fig:Wrinkling_Visualization_Multi}
     \end{subfigure}
     \caption{Different unstable modes for multicomponent vesicle: $(a)$ pearling $(n=0)$ $(b)$ buckling $(n=1)$ $(c)$ wrinkling $(n=2)$, and $(d)$ wrinkling $(n=3)$.}
     \label{fig:three graphs_multi}
\end{figure}

Unlike the single-component system that showed only pearling beyond a particular membrane tension \citep{boedec_jaeger_leonetti_2014,Narsimhan_Spann_Shaqfeh_2015}, multicomponent vesicles can exhibit richer dynamics. The existence of phase separation, line tension, and bending rigidity inhomogeneities can give rise to a combination of pearling, buckling, or wrinkling modes at zero or positive membrane tension. We visualize the shape of some of these modes in figure \ref{fig:three graphs_multi}. The blue color indicates the cholesterol-rich ordered $L_{o}$ phase whereas the yellow phase indicates the cholesterol-less disordered $L_{d}$ phase.

In the following subsections, we will explore these instabilities in greater detail.  We will choose the following parameters in our simulations.  We will examine equiviscous vesicles $(\lambda = 1$) as experiments typically inspect this value \citep{Yanagisawa2010}.  Unless otherwise noted, we will choose a bending difference parameter $\beta = (\kappa_{lo} - \kappa_{ld})/(\kappa_{l0} + \kappa_{ld}) = 0.5$, since we find that $\beta$ in the range listed in Table \ref{tbl:Dimensionless_parameter_range} does not qualitatively alter results.  We will also choose the Peclet number $1 \leq Pe \leq 10$ since previous experimental studies suggest that coarsening and bending timescales are comparable \citep{Negishi2008,Luo_Maibaum_2020}.  This leaves three degrees of freedom remaining – the non-dimensional surface tension $\Gamma$, the Cahn number $Cn$, and the line tension parameter $\alpha$.  The non-dimensional surface tension $\Gamma$ is positive when the vesicle is stretched, and negative when the vesicle is compressed.  The Cahn number $Cn$ and $\alpha$ are related to the line tension.  Large $Cn$ and small $\alpha$ correspond to high values of line tension, which disfavors phase separation and suppresses short wavelength instabilities. 

The structure of the remaining sections are as follows.  Section \ref{sec:condstability} characterizes which modes are the most dominant {and provides a discussion when mode mixing can be present.  Section \ref{sec:GrowthRatesPBW} quantifies the most unstable wavenumbers.}  Section \ref{sec:experimental} performs a qualitative comparison to experiments, while Section \ref{sec:EnAnalysis} performs an energy analysis to understand the mechanism of these instabilities.  Lastly, we make a side note for the special case of $Pe \ll 1$, where analytical solutions to the eigenvalues and eigenvectors are available.  While we believe this case is not physically relevant (see Table \ref{tbl:Dimensionless_parameter_range}), Appendix \ref{App:LowPe} provides details of this analysis for those who are interested.

\subsection{Which modes are most dominant?}\label{sec:condstability}

Here, we delineate the conditions under which the axisymmetric $(n=0)$ instability has the largest growth rate, and the conditions under which the non-axisymmetric instabilities ($n \geq 1$) have the largest growth rate.  We will examine the $n = 0,1,2,3$ modes here since we find that $n > 3$ does not dominate for the parameter ranges simulated. When calculating the most dangerous mode, we explore the wavenumber range $0 < k < 3$.

\begin{figure*}
        \centering
        \begin{subfigure}[b]{0.475\textwidth}
            \centering
            \includegraphics[width=\textwidth]{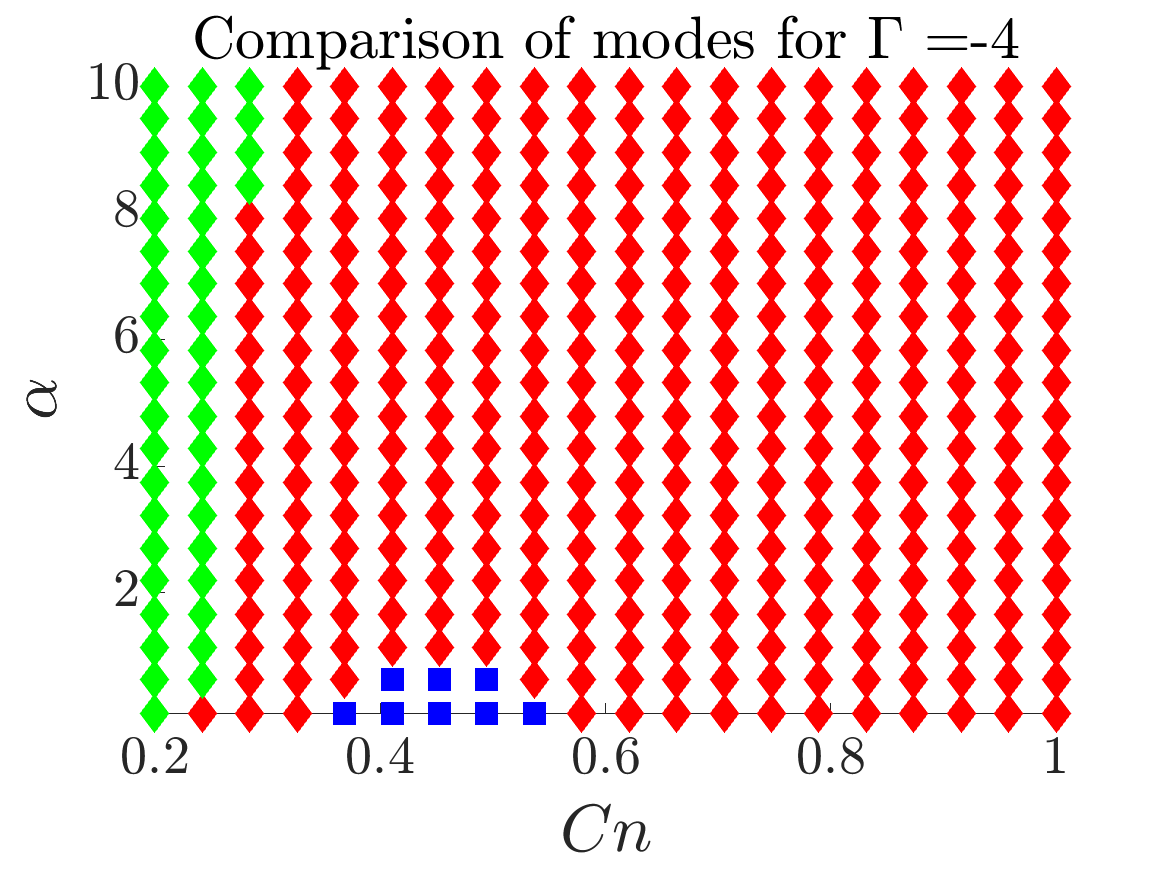}
            \caption[]%
            {{}}    
            \label{fig:ModeDomGammaneg4}
        \end{subfigure}
        \hfill
        \begin{subfigure}[b]{0.475\textwidth}  
            \centering 
            \includegraphics[width=\textwidth]{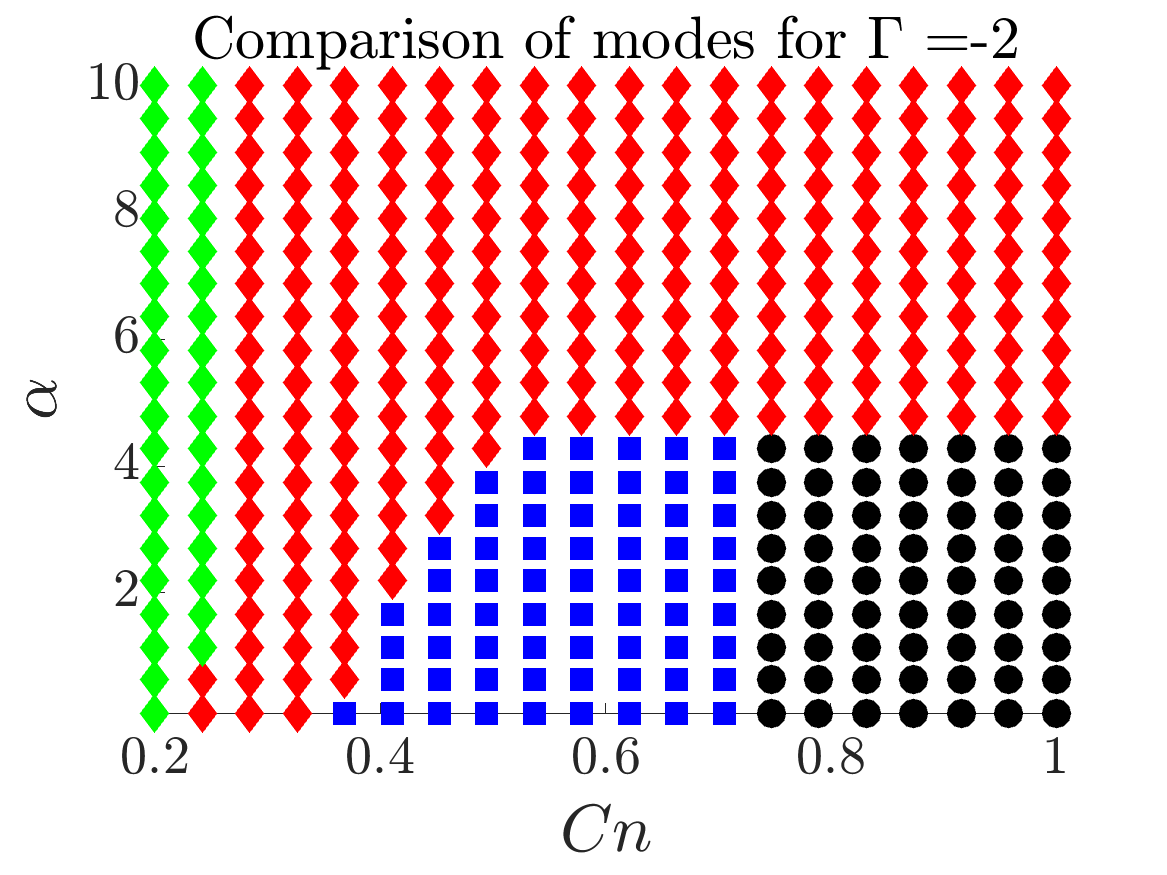}
            \caption[]%
            {{}}    
            \label{fig:ModeDomGammaneg2}
        \end{subfigure}
        \vskip\baselineskip
        \begin{subfigure}[b]{0.475\textwidth}   
            \centering 
            \includegraphics[width=\textwidth]{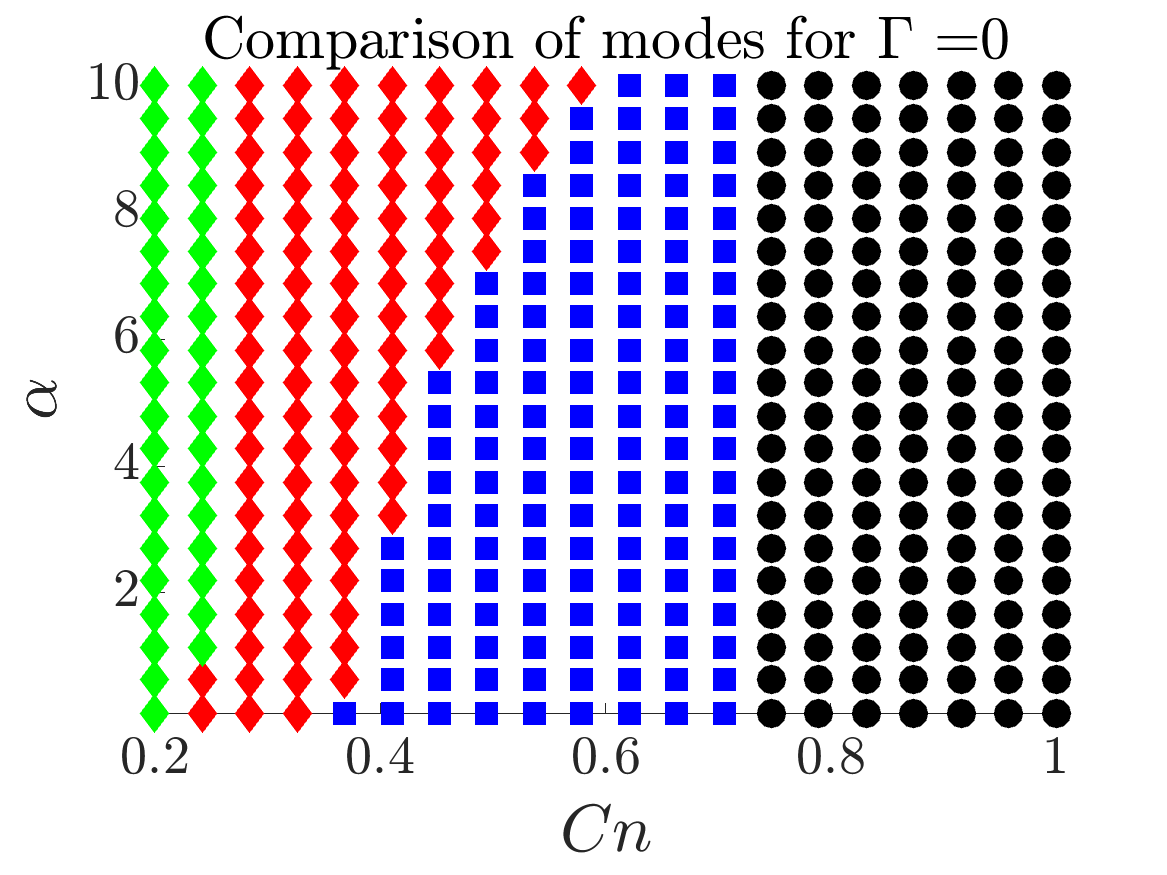}
            \caption[]%
            {{}}    
            \label{fig:ModeDomGamma0}
        \end{subfigure}
        \hfill
        \begin{subfigure}[b]{0.475\textwidth}   
            \centering 
            \includegraphics[width=\textwidth]{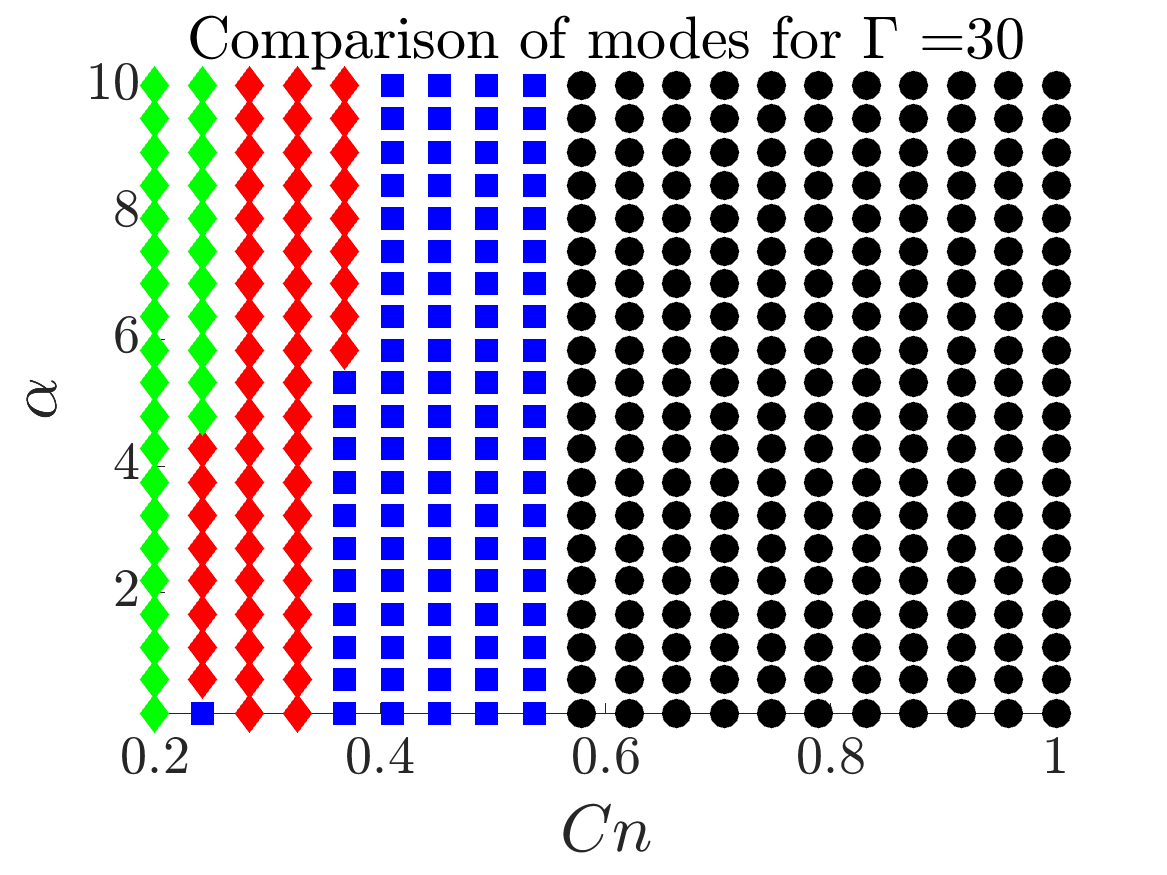}
            \caption[]%
            {{}}    
            \label{fig:ModeDomGamma30}
        \end{subfigure}
        \caption[ The average and standard deviation of critical parameters ]
        {Phase plots for most dominant mode. The black circles represent the case where $n=0$ dominates, the blue squares where $n=1$ dominates, the red diamonds where $n=2$ dominates, and the green diamonds where $n=3$ dominates.  The simulation parameters are $\lambda = 1, Pe = 1, \beta = 0.5$.} 
        \label{fig:ModeDomGamma_Variation}
    \end{figure*}

Figure \ref{fig:ModeDomGamma_Variation} plots which mode has the largest growth rate for different values of the non-dimensional surface tension $(\Gamma)$, Cahn number ($Cn$), and line tension parameter ($\alpha$).  Figure \ref{fig:ModeDomGammaneg4} shows results for a highly compressed vesicle ($\Gamma = -4$), Figure \ref{fig:ModeDomGammaneg2} for a moderately compressed vesicle ($\Gamma = -2$), Figure \ref{fig:ModeDomGamma0} for a vesicle under no tension $(\Gamma = 0$), and Figure \ref{fig:ModeDomGamma30} for a vesicle under strong tension $(\Gamma = 30)$.  Under strong compression ($\Gamma = -4$, Figure \ref{fig:ModeDomGammaneg4}), we see that only the non-axisymmetric modes are dominant $(n \neq 0)$.  This observation is similar to what is seen for single-component vesicles, although we note that for this value of tension $\Gamma = -4$, only the $n = 1$ and $n = 2$ modes are unstable for the single-component case while $n=2$ and $n=3$ mostly dominate for the multicomponent case.  For very small values of the Cahn number (very low line tension), the dominant modes become more non-axisymmetric, a trend that is seen in all four plots here.

When the vesicle is under moderate compression ($\Gamma = -2$, Figure \ref{fig:ModeDomGammaneg2}) or no compression ($\Gamma = 0$, Figure \ref{fig:ModeDomGamma0}), all modes $n = 0, 1, 2, 3$ can be unstable depending on the value of Cahn number $(Cn)$ and line tension parameter $(\alpha$).  These results are very different than what is seen for single-component vesicles where no modes are unstable at zero tension ($\Gamma = 0$) and only the $n=1$ mode is unstable at moderate compression ($\Gamma = -2$).  It also appears that $\alpha$ plays a more significant role in the mode selection than the highly compressed vesicle case ($\Gamma = -4$, Figure \ref{fig:ModeDomGammaneg4}).  

When the vesicle is under large tension ($\Gamma = 20$, Figure \ref{fig:ModeDomGamma30}), the phase plot looks similar to the zero-tension case, except that a larger portion of the phase space shows axisymmetric modes ($n = 0$) being dominant.  When the tension becomes very large $(\Gamma \rightarrow \infty)$, one will only observe pearling modes, recovering the results from the single-component case.

We note that while this analysis shows phase plots for the most unstable modes, it does not comment on the magnitude of these growth rates compared to other modes.  Below, we will see that in many situations, the growth rates of different modes can be comparable and give rise to mode mixing.


\begin{figure}
\centering

\begin{subfigure}{0.49\textwidth}
\includegraphics[width=\textwidth]{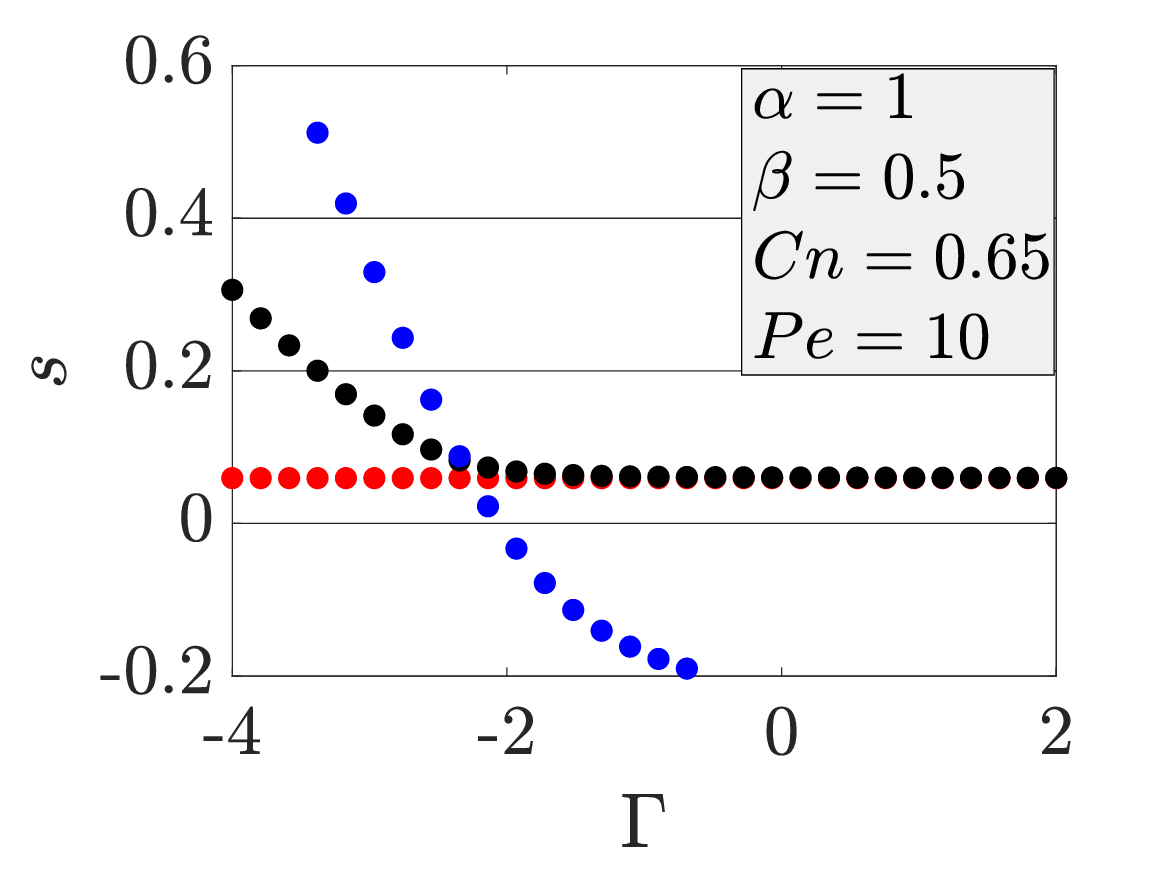}
\captionsetup{font=normalsize,labelfont={bf,sf}}
    \caption{$Cn=0.65$}
    \label{Cn_03_MU_gammadependence}
\end{subfigure}
\begin{subfigure}{0.49\textwidth}
\includegraphics[width=\textwidth]{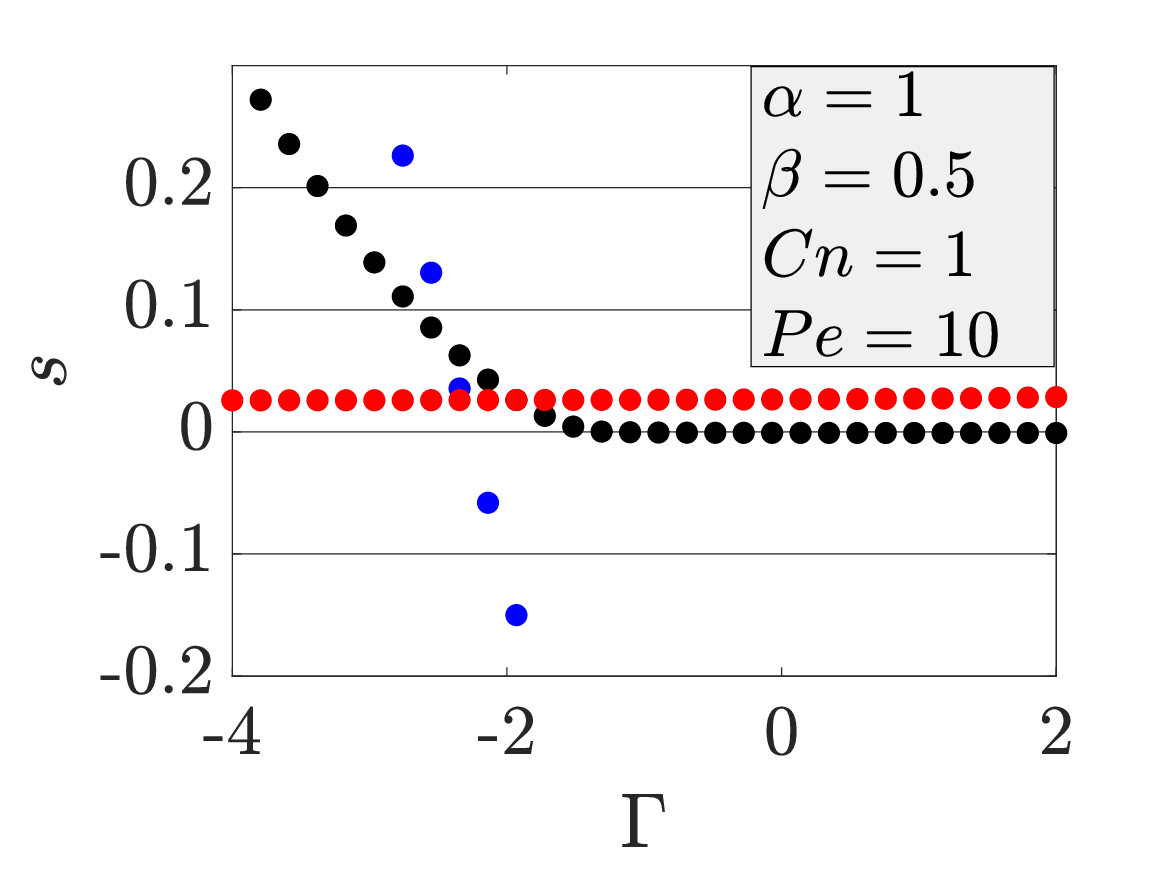}
\captionsetup{font=normalsize,labelfont={bf,sf}}
    \caption{$Cn=1$}
    \label{Cn_1_MU_gammadependence}
\end{subfigure}

\caption{Most unstable growth rates with respect to the isotropic membrane tension $\Gamma$ for multicomponent vesicles. The red circles represent $n=0$ pearling modes, black circles represent $n=1$ buckling modes, and blue circles represent $n=2$ wrinkling modes.  The dimensionless parameters are (a) $\lambda =1, Pe=10, \alpha = 1, \beta = 0.5, Cn =0.65$ and (b) $\lambda =1, Pe=10, \alpha = 1, \beta = 0.5, Cn =1$. }
\label{fig:Most_Unstable_wrt_Gamma_Multi}
\end{figure}

Figure \ref{fig:Most_Unstable_wrt_Gamma_Multi} presents the magnitude of the most unstable growth rates for the three modes $n=0,1,2$ with respect to the isotropic membrane tension $\Gamma$. The dimensionless parameters $\lambda = 1$, $\alpha = 1$, $\beta = 0.5$ and $Pe=10$ are chosen to be representative of experimental values in \citet{Yanagisawa2010} (see \ref{sec:experimental} for more details). Based on the interface width between the ordered and disordered phases, we could have {different values of the Cahn number. We pick two values here: $Cn=0.65,1$}. We observe that for lower Cahn numbers $(Cn = 0.65)$, the buckling and wrinkling modes dominate over pearling modes at compressive values of membrane tension. As the tension increases, the growth rates become comparable for pearling and buckling. As the Cahn number increases to 1, the wrinkling and buckling modes dominate for highly compressive tensions ($\Gamma<-2$) but become stabilized for small compressive and positive values of $\Gamma$ where the pearling modes become dominant. This leads to pure pearling instabilities that will be discussed in detail in section \ref{sec:experimental}.



\subsection{{Wavenumber dependence of growth rates} }\label{sec:GrowthRatesPBW}
\begin{figure}
\centering

\begin{subfigure}{0.49\textwidth}
\includegraphics[width=\textwidth]{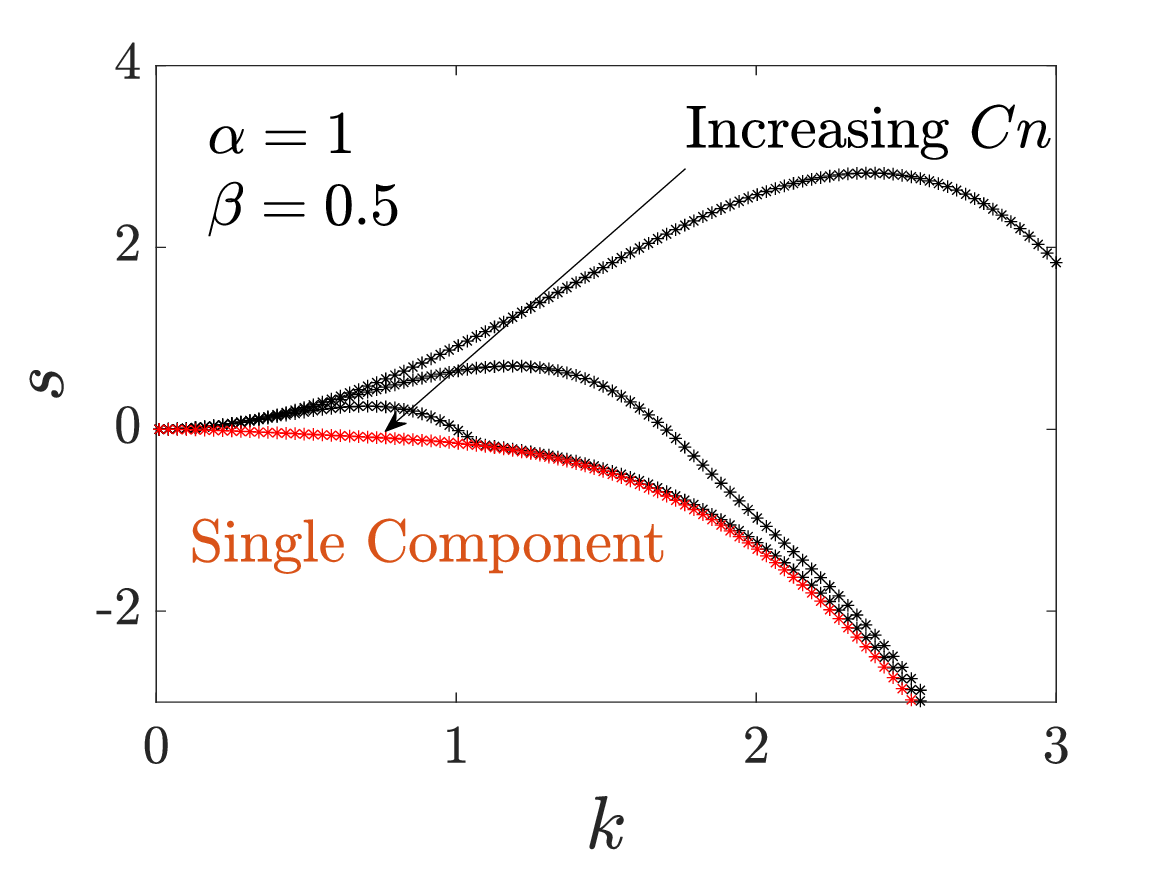}
\captionsetup{font=normalsize,labelfont={bf,sf}}
    \caption{Variation with $Cn$}
    \label{PearlingGrowth_Cn}
\end{subfigure}
\begin{subfigure}{0.49\textwidth}
\includegraphics[width=\textwidth]{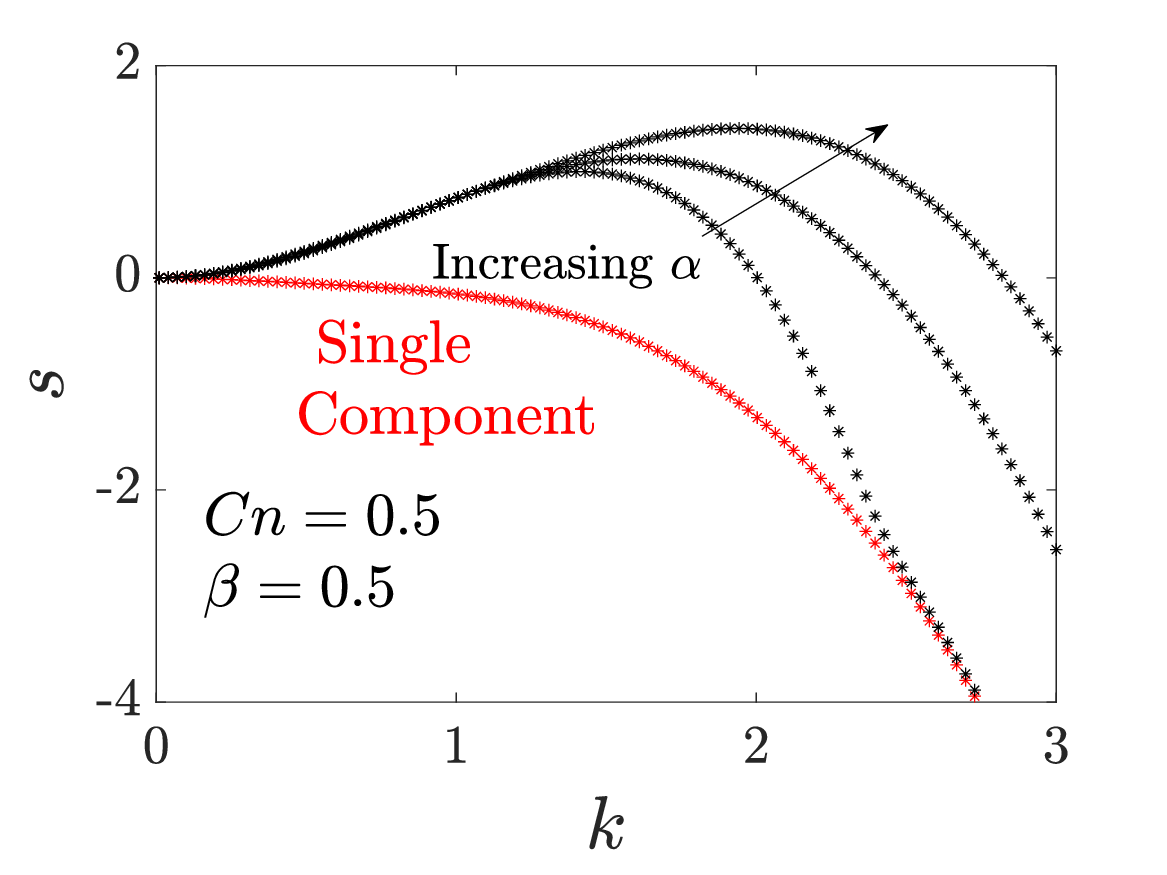}
\captionsetup{font=normalsize,labelfont={bf,sf}}
    \caption{Variation with $\alpha$}
    \label{PearlingGrowth_alpha}
\end{subfigure}

\caption{Growth rate $(s)$ vs wavenumber ($k$) for pearling $(n = 0)$ mode. (a) Dependence on Cahn number ($Cn=0.3,0.6,1$) for $\alpha=1, \beta = 0.5, Pe = 1$.  (b) Dependence on line tension parameter ($\alpha =0.1,10,20$) for $Cn = 0.5, \beta = 0.5, Pe = 1$.  In both graphs, the multicomponent (black) results are compared against single-component (red) results for $\Gamma = 0, \lambda = 1$.}
\label{fig:Pe1_Pearling_GrowthRates}
\end{figure}

\begin{figure}
\centering
\begin{subfigure}{0.49\textwidth}
    \includegraphics[width=\textwidth]{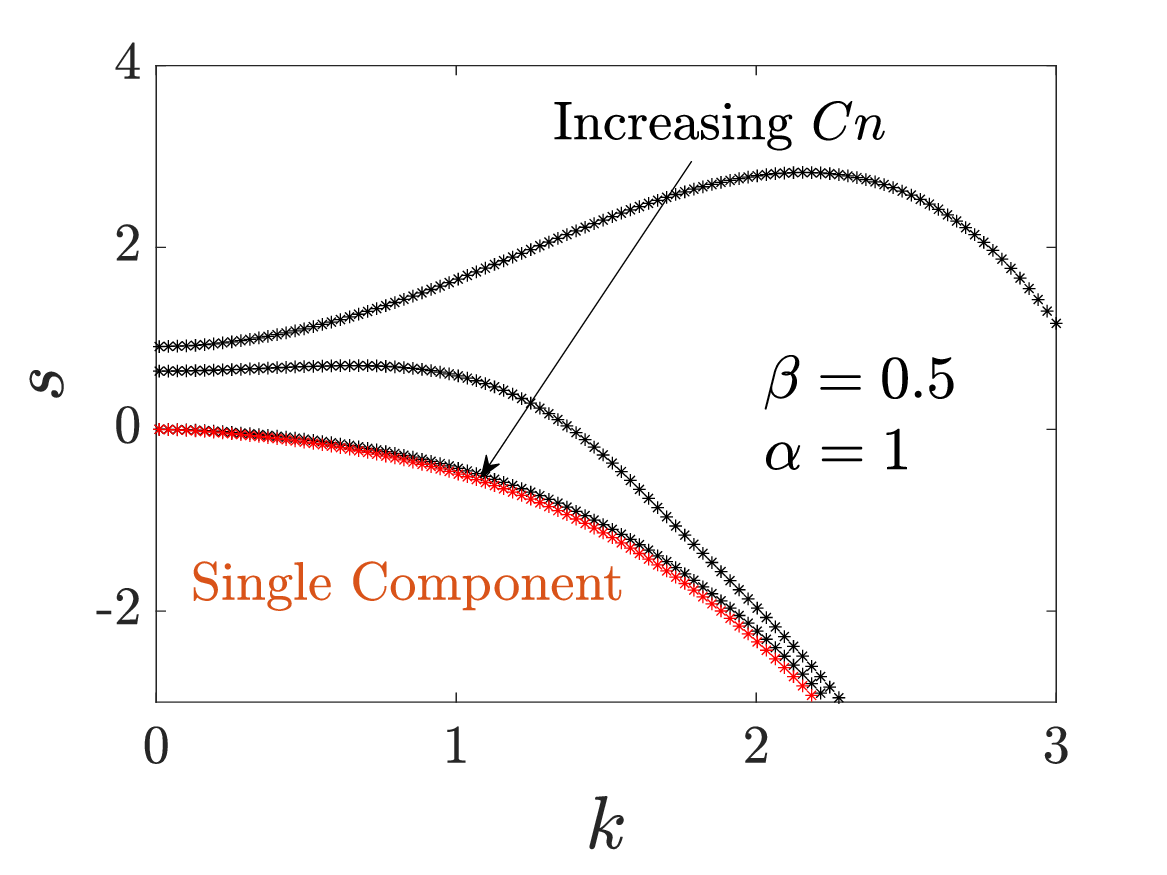}
    \captionsetup{font=normalsize,labelfont={bf,sf}}    
    \caption{Variation with $Cn$}
    \label{BucklingGrowth_Cn}
\end{subfigure}
\begin{subfigure}{0.49\textwidth}
    \includegraphics[width=\textwidth]{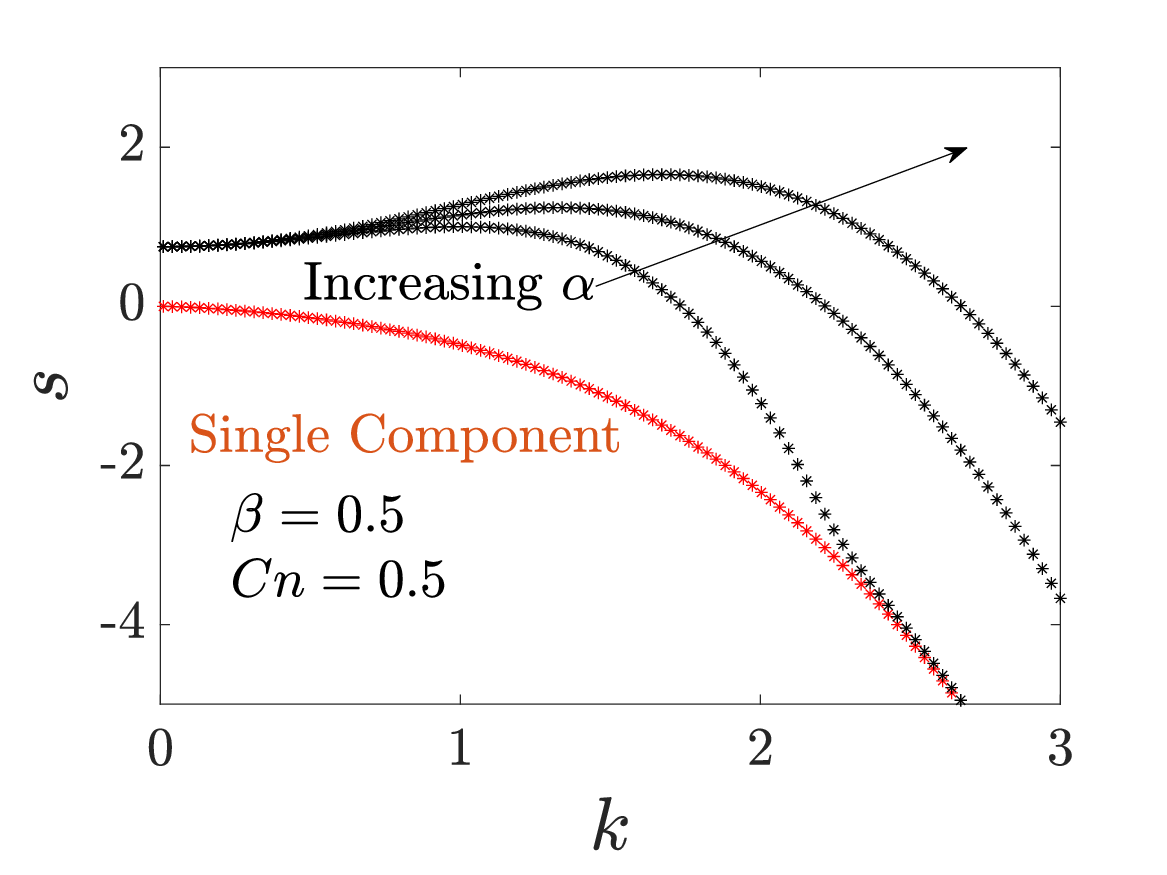}
    \captionsetup{font=normalsize,labelfont={bf,sf}}    
    \caption{Variation with $\alpha$}
    \label{BucklingGrowth_alpha}
\end{subfigure}

\caption{Growth rate $(s)$ vs wavenumber ($k$) for buckling $(n = 1)$ mode. (a) Dependence on Cahn number ($Cn=0.3,0.6,1$) for $\alpha=1, \beta = 0.5, Pe = 1$.  (b) Dependence on line tension parameter ($\alpha =0.1,10,20$) for $Cn = 0.5, \beta = 0.5, Pe = 1$.  In both graphs, the multicomponent (black) results are compared against single-component (red) results for $\Gamma = 0, \lambda = 1$.}
\label{fig:Pe1_Buckling_GrowthRates}
\end{figure}

\begin{figure}
\centering
\begin{subfigure}{0.49\textwidth}    \includegraphics[width=\textwidth]{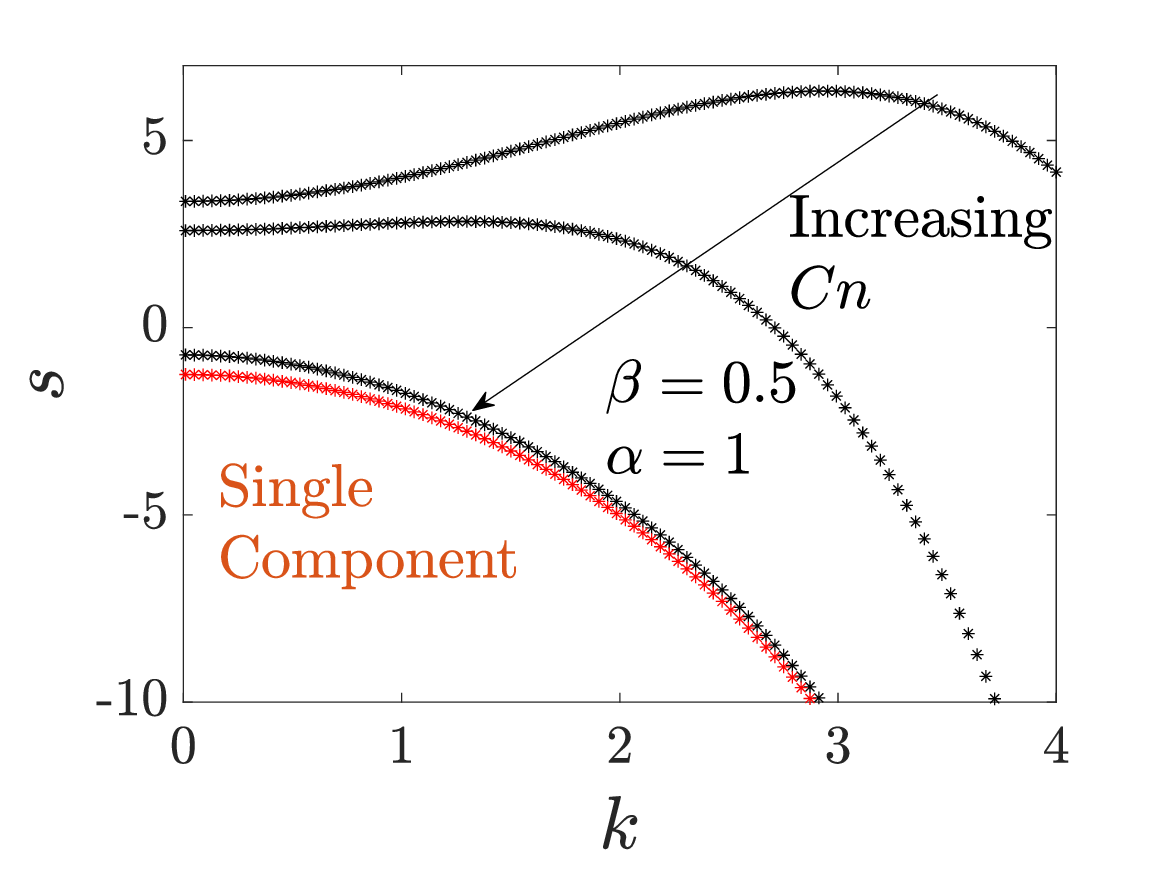}
\captionsetup{font=normalsize,labelfont={bf,sf}}    
    \caption{Variation with $Cn$}
    \label{WrinklingGrowth_Cn}

\end{subfigure}
\begin{subfigure}{0.48\textwidth}
    \includegraphics[width=\textwidth]{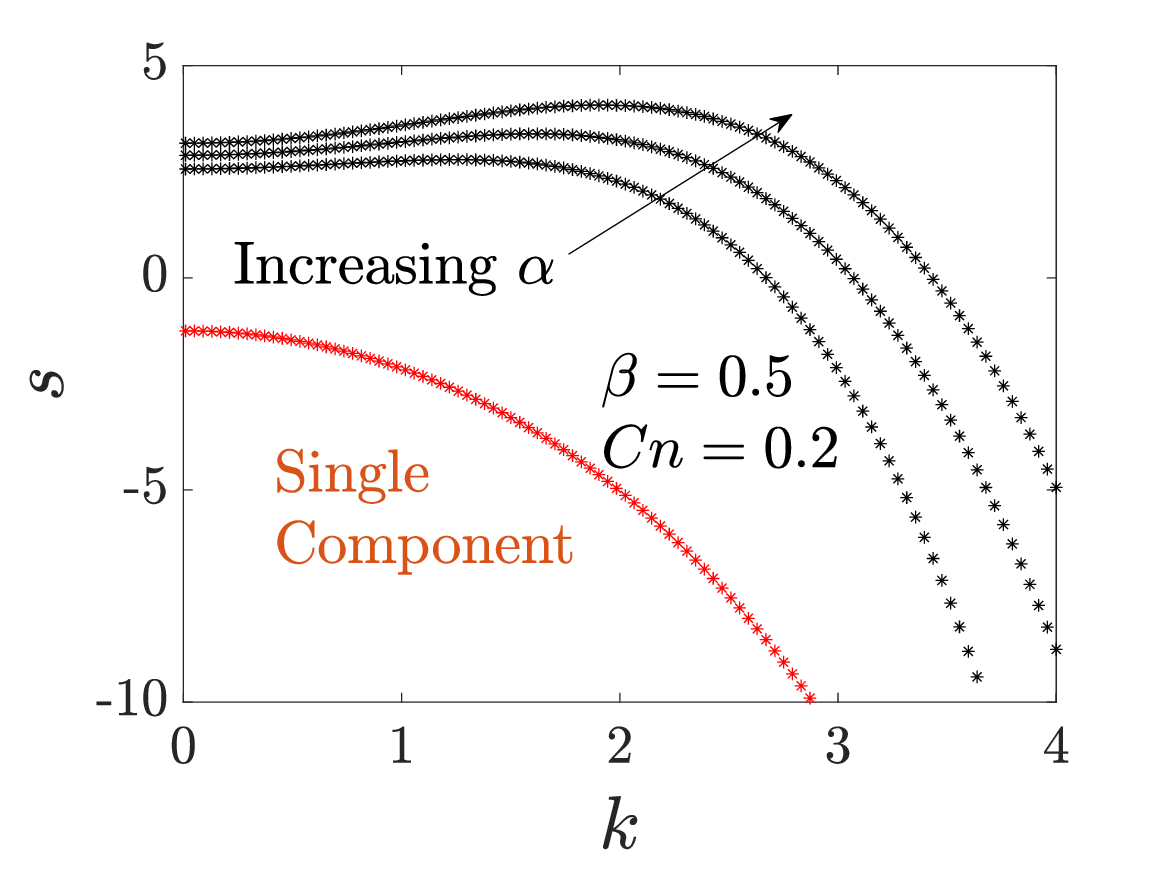}
\captionsetup{font=normalsize,labelfont={bf,sf}}    
    \caption{Variation with $\alpha$}
    \label{WrinklingGrowth_alpha}
\end{subfigure}

\caption{ Growth rate $(s)$ vs wavenumber ($k$) for wrinkling $(n = 2)$ mode. (a) Dependence on Cahn number ($Cn=0.2, 0.3, 0.6$) for $\alpha = 1, \beta = 0.5, Pe = 1$.  (b) Dependence on line tension parameter ($\alpha =0.1, 10, 20$) for $Cn = 0.2, \beta = 0.5, Pe = 1$.  In both graphs, the multicomponent (black) results are compared against single-component (red) results for $\Gamma = 0, \lambda = 1$.}
\label{fig:Pe1_Wrinkling_GrowthRates}
\end{figure}

{Figures \ref{fig:Pe1_Pearling_GrowthRates}-\ref{fig:Pe1_Wrinkling_GrowthRates} plot the wavenumber dependence of the growth rates for different instabilites -- the pearling mode ($n = 0$, figure \ref{fig:Pe1_Pearling_GrowthRates}), buckling mode ($n = 1$, figure \ref{fig:Pe1_Buckling_GrowthRates}), and wrinkling mode ($n=2$,  figure \ref{fig:Pe1_Wrinkling_GrowthRates}).  In these plots, the membrane tension is $\Gamma=0$.  Generally, we observe the following trends:  as the Cahn number $Cn$ increases and the line tension parameter $\alpha$ decreases, the maximum growth rate decreases and the most dangerous wavenumber decreases (i.e., the wavenumber $k$ corresponding to maximum growth rate).  These trends occur because large $Cn$ and small $\alpha$ values correspond to large line tensions, which suppresses growth rates and disfavors short wavelength (i.e., large $k$) instabilities.  We note that the extent to which the growth rates are altered depends greatly on the mode number $(n)$ -- this is why for certain values of $(\alpha, Cn)$, the pearling modes have the largest growth rate, but for other values the non-axisymmetric modes have the largest growth rate.  We also see that while large $Cn$ and small $\alpha$ values suppress short wavelength (i.e., $k > 1$) instabilities, $Cn$ plays a more significant role in altering the low wavenumber ($k < 1$) growth rates compared to $\alpha$.  }

{Some of our trends seem consistent with previous simulations of non-tubular vesicles \citep{GeraThesis}. Specifically, the cited study found that increasing $\alpha$ forms shorter wavelength (larger $k$) stripes on the vesicle, consistent with our study.  However, Gera finds that as $\alpha$ rises, it appears that the time slows down to reach the observed behaviour, which is opposite of the growth rate trends observed here (see figures \ref{PearlingGrowth_alpha} - \ref{WrinklingGrowth_alpha}). 
 We point the reader to several caveats:  first, the study by Gera inspects non-tubular vesicles, which is different than the geometry considered here.  Secondly, the study examines the full nonlinear dynamics, whereas we inspect the linearized dynamics and hence the onset of instabilities.  We cannot ensure that these instabilities will persist during longer time durations. This analysis is left for a later study.}


Lastly, we inspect the variation of the most unstable wavenumber with respect to the membrane tension $\Gamma$ for different modes $n=0,1,2$ at the experimentally realizable ranges of parameters. Since this variation is not large, we have added these plots to the appendix (see Appendix \ref{app:wavenumber_Dependence}).

\subsection{Experimental comparison}\label{sec:experimental}

In this section, we compare the instabilities from our linear stability analysis to experimental observations from \citep{Yanagisawa2010}.  In this paper, the authors explored periodic modulations in cylindrical, multicomponent vesicles containing DOPC/DPPC/cholesterol at 1:1 DOPC:DPPC and different amounts of cholesterol. The vesicles were created by taking spherical giant unilamellar vesicles (GUVs) with these lipids, and osmotically deflating them to create tubular shapes of radius $R \approx 0.5-3 \mu m$ with aspect ratios between $L = 5-20$.  The modulations observed arose due to the phase separation into liquid-ordered $(L_o)$ and liquid-disordered $(L_d)$ phases, similar to what is seen in our theories. The interior and exterior fluids were the same (up to the sugars used for osmotic deflation), yielding a viscosity ratio $\lambda \approx 1$. {Based on the ratios of DOPC:DPPC:chol in their studies, the average bending stiffness was estimated to be $k_0 \approx 10^{-19} J$ and the difference in bending stiffness between domains varied between $\beta = 0.1 – 0.5$.  The line tension was estimated to be roughly $1 pN$, yielding a line tension parameter $\alpha \approx 1$.  Examining the interface width yields a Cahn number $Cn = \varepsilon^{width}/(\sqrt{2}R) \approx 0.3-1$.  We find that results are highly sensitive to this parameter as shown below.   The surface Peclet number was estimated to be $Pe = O(1)$ based on limited data of lipid diffusivities \citep{Negishi2008}.}

The only non-dimensional number we were not able to infer from experimental data was the dimensionless surface tension $\Gamma = \sigma_0 R^2/k_0$, since the surface tension $\sigma_0$ was not provided.  {In principle, one could obtain $\sigma_0$ by performing an equilibrium simulation of vesicle shape since this quantity arises as a Lagrange multiplier that enforces the constant area of the membrane.}  However, this simulation is quite difficult to do for highly deflated, multicomponent vesicles (and to our knowledge has yet to be performed).  {Instead, we make a note that $\sigma_0$ is likely to be very small since the vesicles are under no external force, and for values $\sigma_0 = 10^{-7} N/m$, this yields $\Gamma \sim O(1)$. Thus, we will perform simulations for several different values of $\Gamma$ and see how they compare against experimental data.  We will also vary $Cn$ since the results are sensitive to this value.  For the other parameters, we set $\lambda = 1, Pe =10, \alpha = 1, \beta = 0.5$ consistent with estimated values listed above.}



\begin{figure}
\centering

\includegraphics[width=0.2\textwidth]{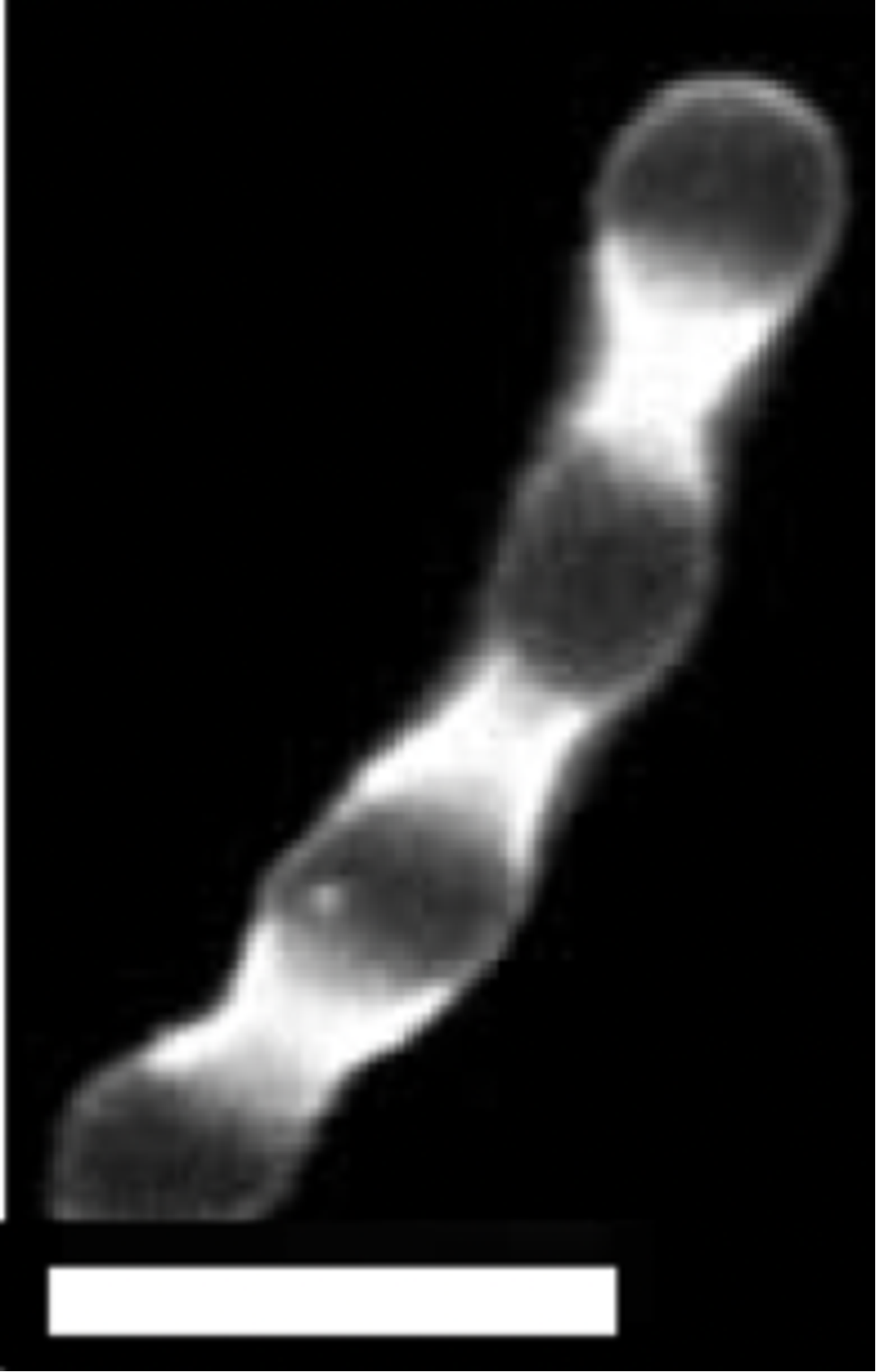}
\includegraphics[width=0.48\textwidth]{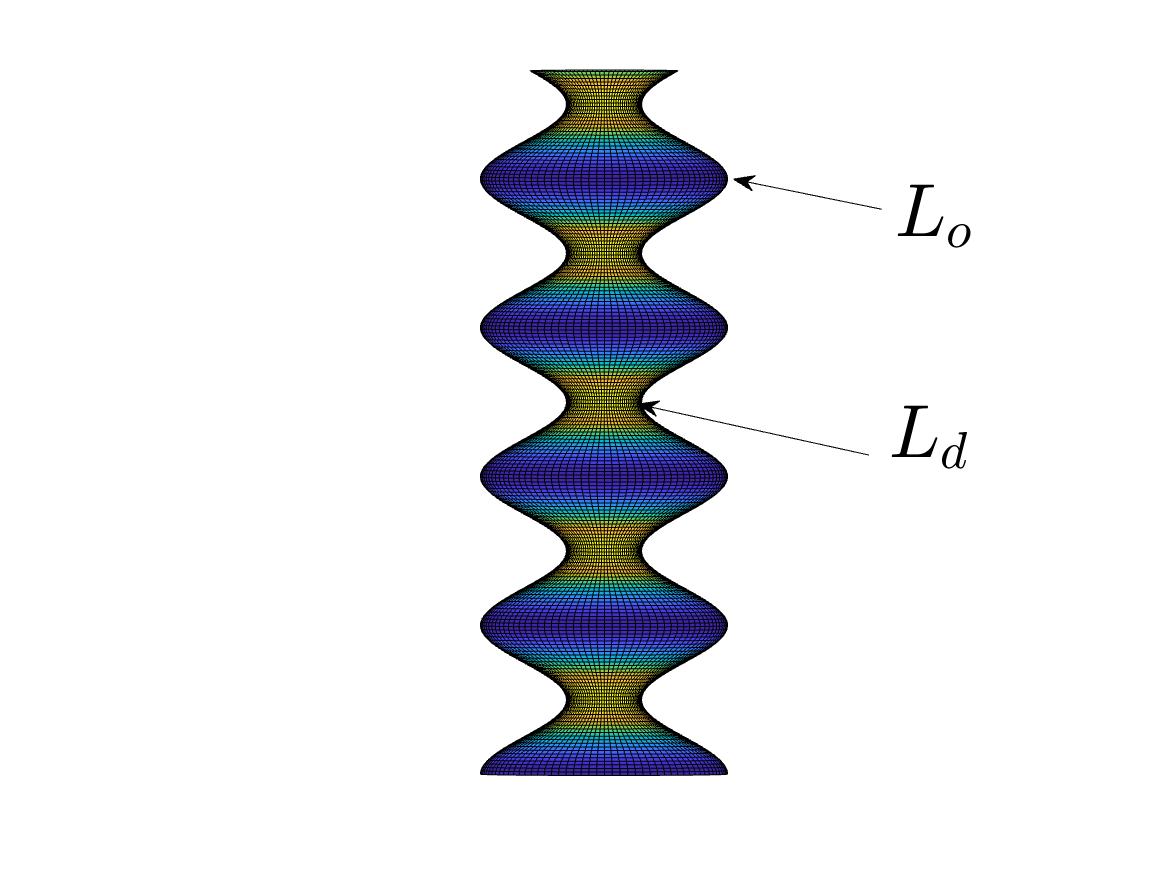}
\caption{Pearling visual qualitative comparison with experiments \citep{Yanagisawa2010} where the blue domains represent the cholesterol rich $L_{o}$ phase (black in experiments) and the yellow domains (white in experiments) represent the cholesterol-lacking $L_{d}$ phase. The parameters for the simulation are $\lambda = 1, Pe =10, \alpha = 1, \beta = 0.5, Cn =1$ and $\Gamma = 0$ corresponding to a vesicle radius $R\approx 1 \mu m$.  The initial condition is the most unstable pearling mode with $L_2$-norm 0.001, and the results are {simulated up to time $t =300$, which translates to an experimental time of $\approx 3$ seconds. The scale bar represents a length of $5 \mu m$. The mole fraction ratio of DOPC:DPPC:Chol is 9:9:22.} }
\label{fig:Pearling_Comparison_Exp}
\end{figure}

In Figure \ref{fig:Pearling_Comparison_Exp}, we show one snapshot of an experimental image where the vesicle forms a straight line with pearls.  The {bright regions} represent the disordered $L_{d}$ phase and the {dark regions} represent the ordered $L_{o}$ phase.  The interface width is fairly diffuse, leading to $Cn \approx 1$.  If we perform simulations with no tension $\Gamma = 0$, we observe qualitatively similar behaviour to experiments.  We observe that the axisymmetric, pealing mode ($n = 0$) is the dominant instability ($s=0.0266$), while the other modes are stable. We also note that the stiff and soft lipids accumulate in the peaks and troughs of the profile, respectively.  The wavelength is longer than the radius, indicating $k < 1$ ($k \approx 0.701$), although in order to match the experiments quantitatively, one will have to tune the membrane tension. A tension $\Gamma > 0$ yields qualitatively the same results.

\begin{figure}
\centering

\includegraphics[width=0.25\textwidth]{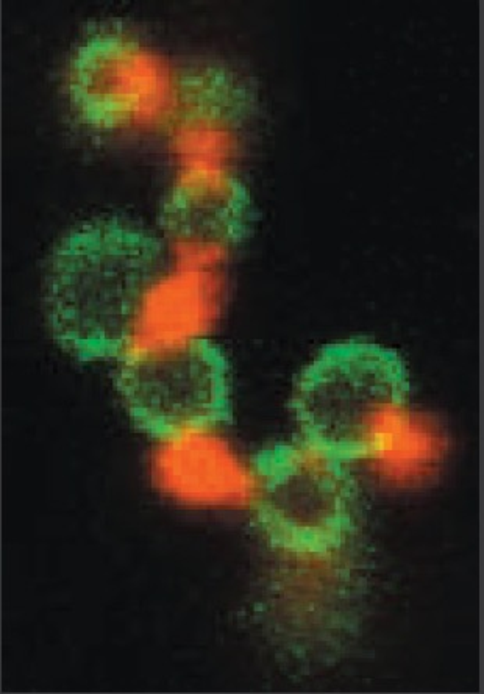}
\quad \includegraphics[width=0.6\textwidth]{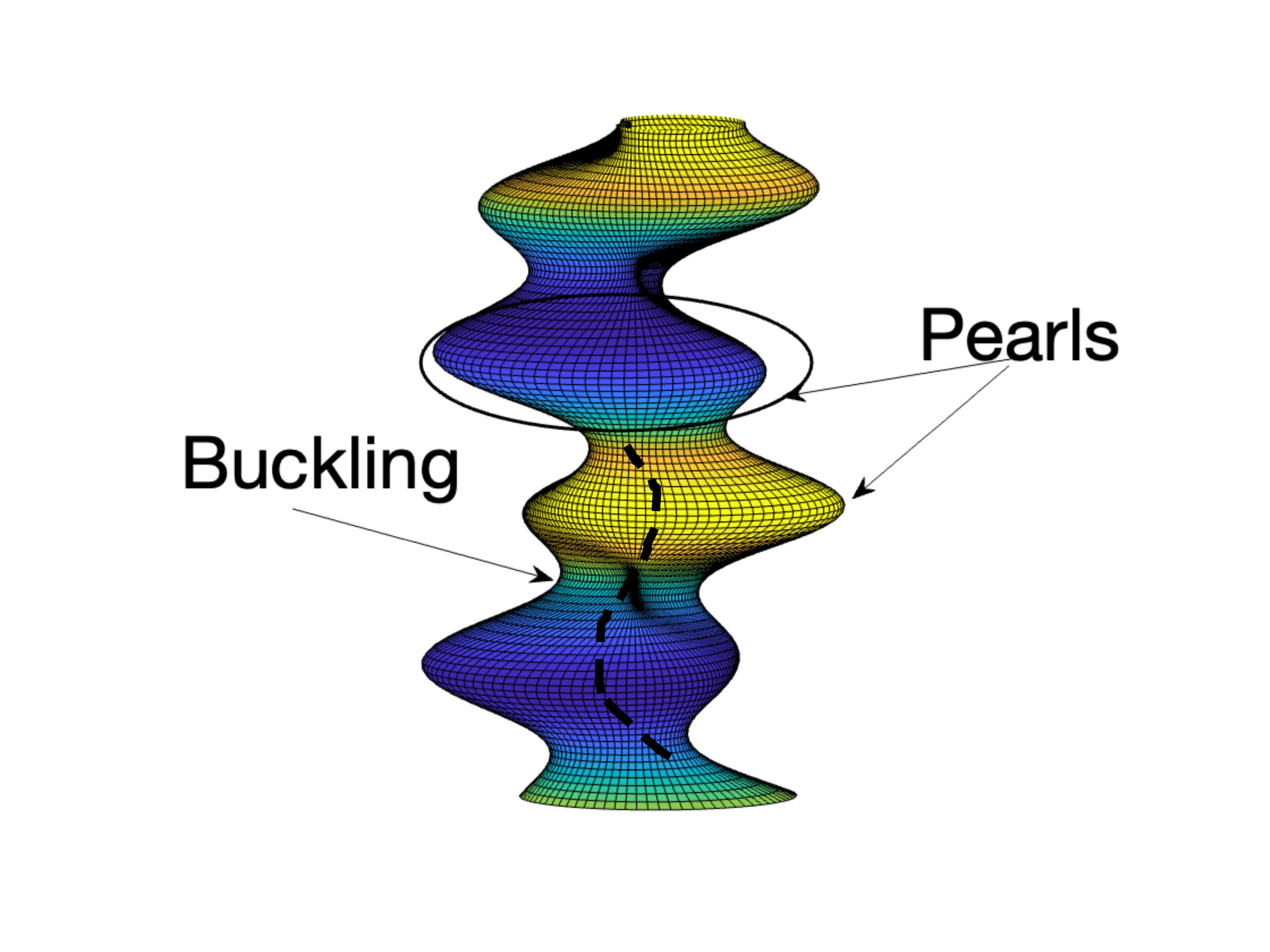}

\caption{Mixed mode instability found in  experiments \citep{Yanagisawa2010} and simulations. The pearling mode $(n = 0)$ can have a larger wavenumber compared to the buckling mode $(n = 1)$.  The parameters for the simulation are $\lambda = 1, Pe = 10, \alpha = 1, \beta = 0.5, Cn=0.65,$ and $\Gamma =2$ corresponding to $R=1\mu m$.   The initial condition is the sum of the most unstable pearling and buckling modes $[r_{kn},q_{kn}]$, with $L_2$-norm 0.004 and 0.001 respectively, and the results are simulated up to time $t = 140$ which translates to a physical time of $\approx 1.4$ seconds. The scale bar represents a length of $2\mu m$. The mole fraction ratio DOPC:DPPC:Chol is 3:3:4. }
\label{fig:Buckling_Comparison_Exp}
\end{figure}

In some of the images observed in the paper, the vesicles exhibited buckling in addition to pearling. In these situations, the interface width appears sharper than the case when pearls form.  Figure \ref{fig:Buckling_Comparison_Exp} shows a snapshot of such an example.  If we perform a simulation with $Cn =0.65$ and slightly positive tension $\Gamma =2$, one finds that while the non-axisymmetric buckling mode ($n = 1$) is technically dominant (growth rate $s = 0.0597$), the axisymmetric pearling mode $(n = 0)$ has a growth rate ($s = 0.0593$) with nearly the same value (the other modes are stable).  The most unstable wavenumbers are $k=1.0939$ for pearling and $k=0.4709$ for buckling.  Superimposing the pearling and buckling modes at time $t = 140$, which translates to a physical time of $t = 140t_{bending} \approx 1.4$ seconds (considering a vesicle of size $R\approx 1\mu m$ from the scale bar in figure \ref{fig:Buckling_Comparison_Exp}), yields the simulation snapshot shown, which captures the same qualitative shape seen in the experiment – e.g., pearling modes having a shorter wavelength the buckling modes.  For these mixed mode instabilities, the shape observed is sensitive to the initial condition in the simulation.


{In short, the interface width and membrane tension could potentially cause buckling modes to jump into the foreground. However, there could be other reasons that could give rise to the observed mixed-mode behaviour.} For example, we found that changing the line tension parameter $\alpha$ to larger values can also achieve the same effect, although based on experimental data we believe this explanation is unlikely to be valid.  We also note that in our theory, we assumed the base state has equal amounts of stiff and soft lipids – i.e., $q_0 = 0$.  The experiments didn’t always follow this 50:50 split, and this could potentially give rise to different shape phenomena. This opens the door for future studies with non-zero average $q_{0}$ values.

\subsection{Energy Analysis}\label{sec:EnAnalysis}

There are three energetic contributions to the instability:  bending energy, phase energy, and surface tension energy (see Section \ref{sec:Membrane_energy}).  To understand which contributions play the largest role, we perform the following analysis.  We take the base state of the cylindrical vesicle ($r_0 = 1, q_0 = 0$) and perturb the radius and concentration as follows:

\begin{equation}
        r = 1 + \epsilon r_{kn} \cos{(kz+n\phi)} - \frac{1}{4}\epsilon^{2} r_{kn}^2  
\end{equation}
\begin{equation}
        q = \epsilon q_{kn} \cos{(kz+n\phi)} - \frac{1}{2} \epsilon^2 q_{kn} r_{kn}  
\end{equation}

The higher order terms are present in order to conserve volume and order parameter to $O(\epsilon^2)$:  i.e., $V = V_0$ and $\int q dS = 0$.  We then compute the change in energy between the perturbed and base states, and break them into the bending ($b$), phase ($p$), and surface tension ($\sigma$) contributions:

\begin{align}
\Delta E &= E(r_{kn}, q_{kn}) – E(r_{kn} = 0, q_{kn} = 0) \\
&= \Delta E_b + \Delta E_p + \Delta E_{\sigma}
\end{align}

If $\Delta E < 0$, the perturbation has a lower energy than the base state, which leads to instability.  If $\Delta E > 0$, the perturbation has a higher energy than the base state, and thus the base state is locally stable.  Below are the bending, phase, and surface tension contributions to the energy change per unit length of the vesicle.  The algebraic details are given in Appendix \ref{app:energy}.

\begin{equation}
    \Delta E_b = \frac{\pi \epsilon^2 {r_{kn}}^2}{2}\left(2k^{2} +{(k^{2}+n^{2})\left(k^{2}+n^{2}-\frac{5}{2}\right) + \frac{3}{2}}\right) + \beta \pi\epsilon^{2} r_{kn} q_{kn} (k^{2}+n^{2}-1)
\end{equation}
\begin{equation}\label{phase_energy}
    \Delta E_p = \frac{\pi \epsilon^2 q_{kn}^2}{2 \alpha Cn^2}\left[\tilde{a} + Cn^2 (n^2 + k^2)\right]
\end{equation}
\begin{equation}\label{tension_energy}
    \Delta E_{\sigma} = \frac{\Gamma\pi\epsilon^2 r_{kn}^2}{2}(n^2 + k^2 -1)
\end{equation}

\begin{figure}
\centering

\begin{subfigure}{0.48\textwidth}    \includegraphics[width=\textwidth]{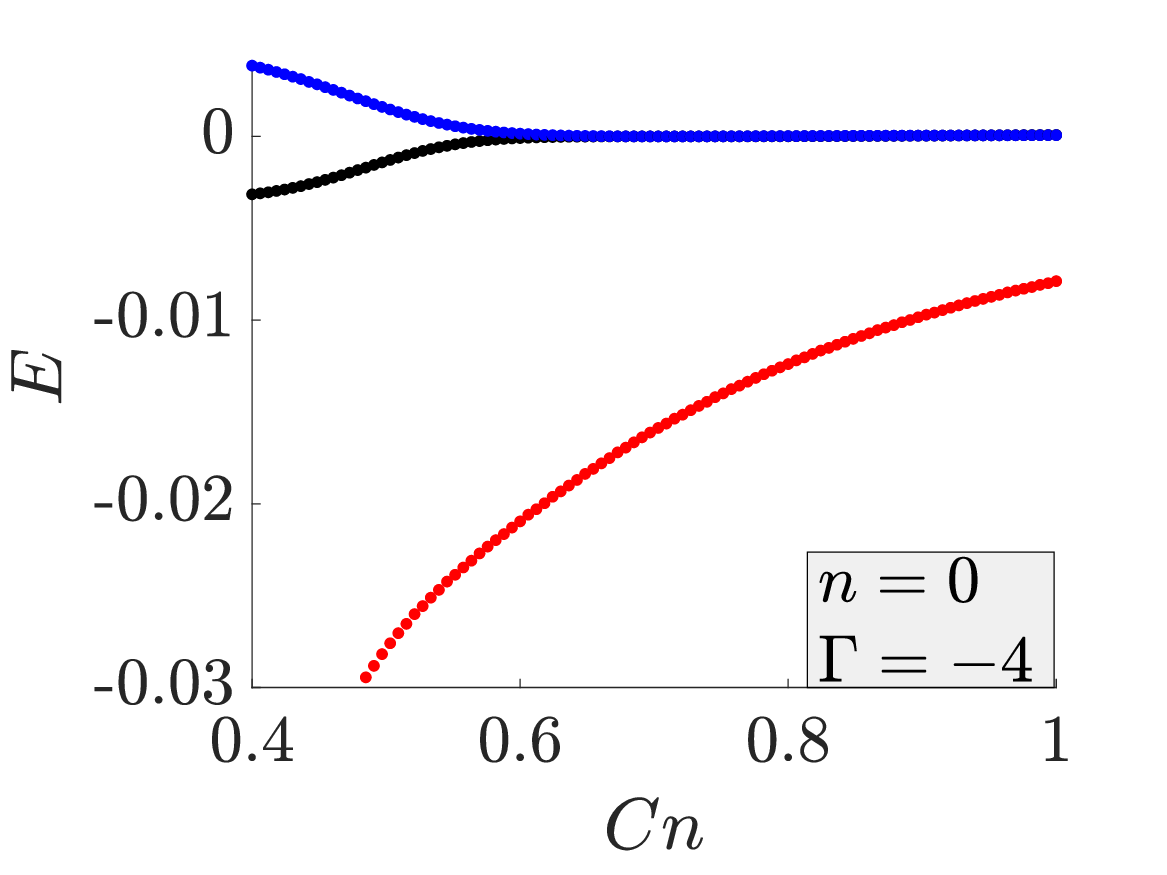}
\captionsetup{font=normalsize,labelfont={bf,sf}}    
    \caption{$\Gamma=-4$}
    \label{WrinklingGrowth_Cn}
    \end{subfigure}
\quad
\begin{subfigure}{0.48\textwidth}    \includegraphics[width=\textwidth]{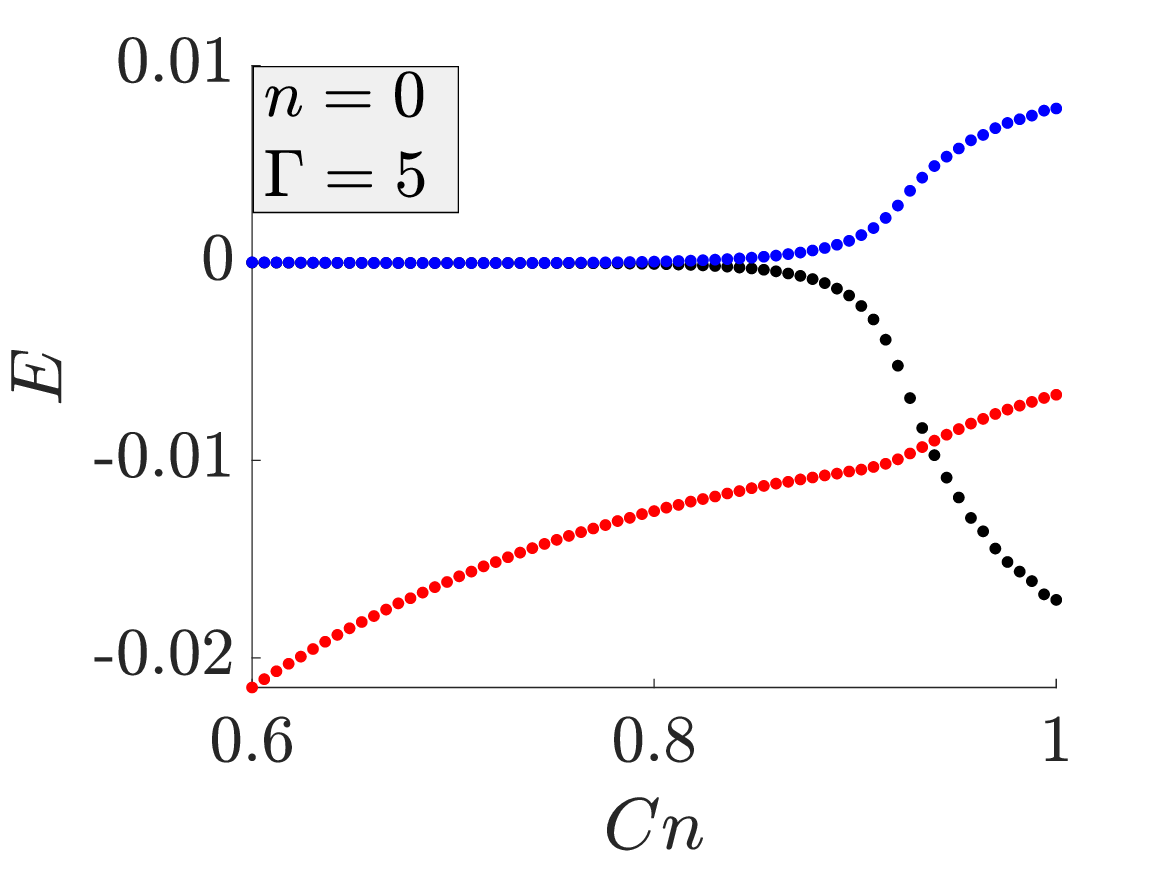}
\captionsetup{font=normalsize,labelfont={bf,sf}}    
    \caption{$\Gamma =5$}
    \label{WrinklingGrowth_Cn}
    \end{subfigure}
\caption{Energetic contributions to the pearling mode ($n=0$) for different values of $Cn$ and $\Gamma$. The red circles represent phase energy ($\Delta E_p$), blue circles represent the bending energy ($\Delta E_b$), and the black circles represent the surface tension energy ($\Delta E_{\sigma}$).  The parameters are $\lambda = 1, Pe = 3, \alpha = 1,\epsilon=0.1$. }
\label{fig:Energy_Comparison_n0}
\end{figure}

\begin{figure}
\centering

\begin{subfigure}{0.48\textwidth}    \includegraphics[width=\textwidth]{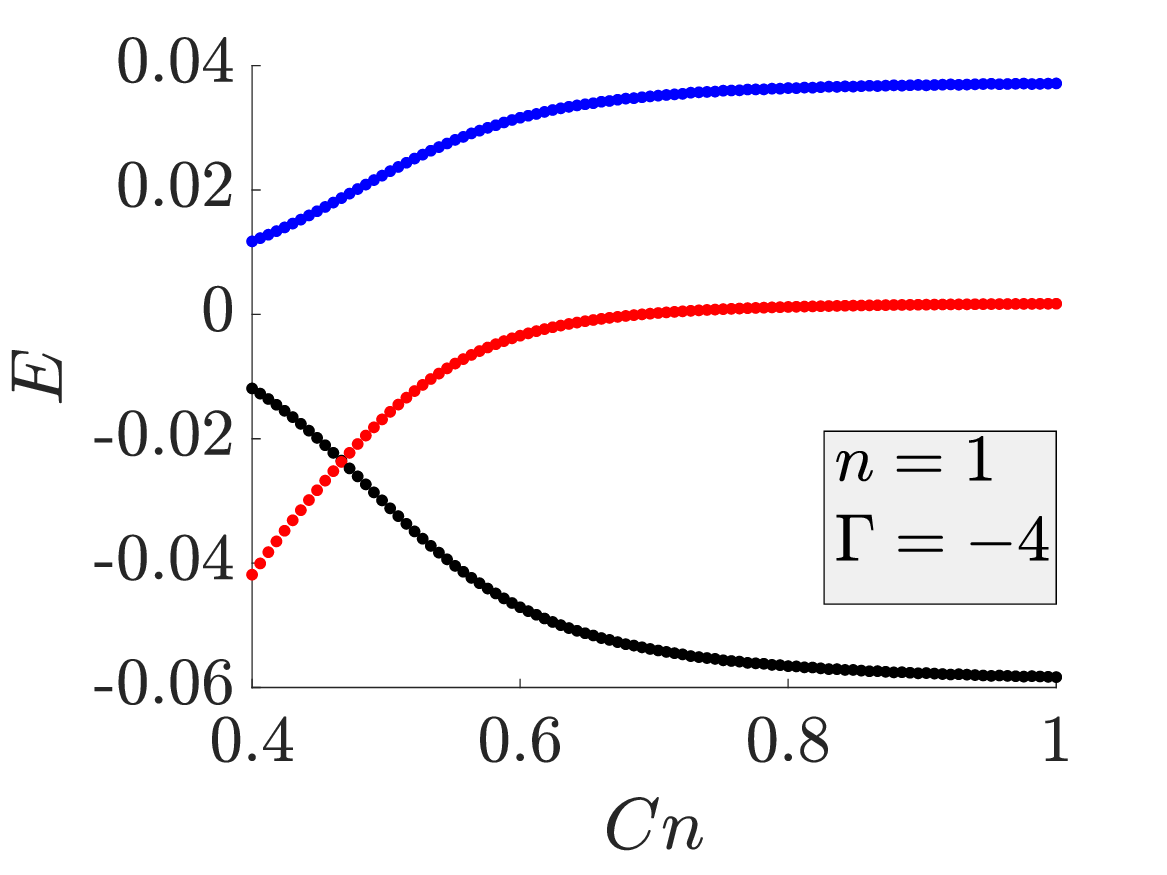}
\captionsetup{font=normalsize,labelfont={bf,sf}}    
    \caption{$\Gamma=-4$}
    \label{E_var_n1_gamma_neg4}
    \end{subfigure}
\quad
\begin{subfigure}{0.48\textwidth}    \includegraphics[width=\textwidth]{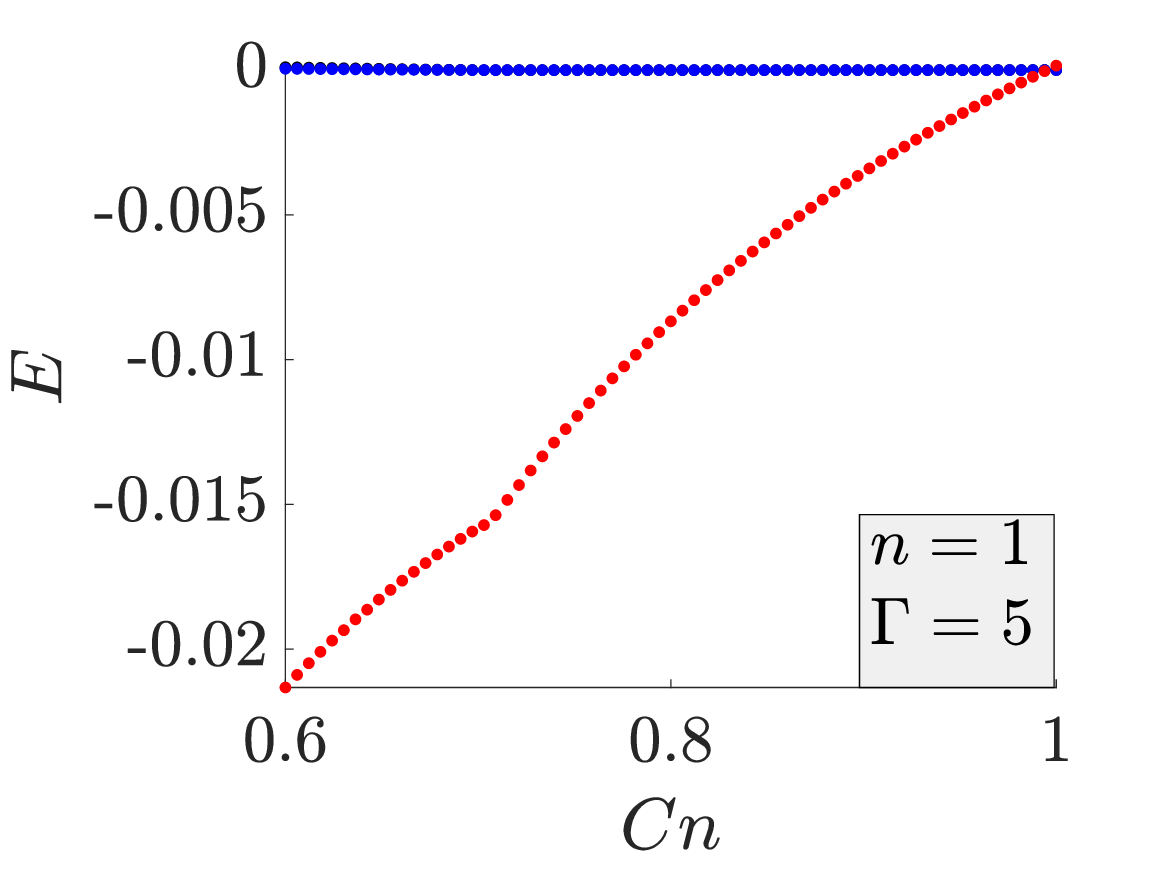}
\captionsetup{font=normalsize,labelfont={bf,sf}}    
    \caption{$\Gamma =5$}
    \label{E_var_n1_gamma_5}
    \end{subfigure}
\caption{Energetic contributions to the buckling mode ($n=1$) for different values of $Cn$ and $\Gamma$. The red circles represent phase energy ($\Delta E_p$), blue circles represent the bending energy ($\Delta E_b$), and the black circles represent the surface tension energy ($\Delta E_{\sigma}$).  The parameters are $\lambda = 1, Pe = 3, \alpha = 1,\epsilon=0.1$. }
\label{fig:Energy_Comparison_n1}
\end{figure}

In figure \ref{fig:Energy_Comparison_n0}, we examine the energetic contributions to the pearling $(n = 0)$ mode at $\lambda=1,Pe=3,\alpha=1$. Here, we use the linear stability theory to compute the dominant eigenvector $[r_{kn}, q_{kn}]$ at the most unstable wavenumber $k$, and then compute the energetic contributions ($\Delta E_b, \Delta E_p, \Delta E_{\sigma}$) as stated above {for perturbation value $\epsilon = 0.1$.  We vary the value of $Cn$ while keeping $\Gamma$ fixed at $-4$ and $5$, both representing extremes of compression and tension respectively.  It can be seen that for experimentally relevant values of the Cahn number $Cn$, the phase energy is the primary driver for the destabilization of the vesicle shape for highly compressive values of $\Gamma$ (figure \ref{fig:Energy_Comparison_n0}a). The bending energy seems to have a stabilizing effect on the vesicle pearling whereas the tension has a weakly destabilizing effect. When the value of $\Gamma$ is largely positive, as the $Cn$ increases, the tension energy begins destabilizing the vesicle more than the phase energy whereas the bending energy is always stabilizing (figure \ref{fig:Energy_Comparison_n0}b).

In figure \ref{fig:Energy_Comparison_n1}, we examine the energetic contributions to the buckling $(n = 1)$ mode at $\lambda=1,Pe=3,\alpha=1$. We vary the value of $Cn$ while keeping $\Gamma$ fixed at $-4$ and $5$, both representing extremes of compression and tension respectively.  It can be seen that for experimentally relevant values of the Cahn number $Cn$, for highly compressive values of $\Gamma$, the tension energy causes the largest destabilization of the vesicle shape as the $Cn$ increases whereas the phase energy contributes less (figure \ref{fig:Energy_Comparison_n1}a). The bending energy seems to have a stabilizing effect on the buckling. When the value of $\Gamma$ is largely positive (figure \ref{fig:Energy_Comparison_n1}b), we see that the phase energy is the primary driver for the destabilization of the vesicle shape.

\section{Conclusions}\label{sec:conclusions}
We performed a linear stability analysis on a tubular vesicle containing multiple components in its bilayer structure.  We observed that the vesicle could exhibit pearling, buckling, and wrinkling behaviour even in the absence of any membrane (surface) tension $\Gamma$, a result that is not seen in single-component vesicles. We determined the conditions under which axisymmetric and non-axisymmetric modes experience the largest growth rate, as well as characterized the growth rates and the wavenumber selection for each mode.  Interestingly, in many situations the axisymmetric pearling mode $(n=0)$ can have similar growth as a buckling mode $(n=1)$, giving rise to a mixed mode instability.  We compared our results to experiments and were able to qualitatively capture similar shape and phase separation patterns \citep{Yanagisawa2010}.  We provided an energy phase diagram to explain the driving forces behind this instability. We saw that there is an interplay between the bending energy, phase energy, and the membrane tension energy, and the dominant contribution depends on the surface tension, line tension, and bending moduli of the domains.

This study brings to light the importance of understanding flow dynamics being coupled with line tension and bending inhomogeneity effects, which opens up a large phase space to be studied. We also note that while the thermodynamic model (Ginzburg-Landau) helps us qualitatively understand some physical phenomena, a detailed use of more complicated models and their dependence on membrane tension and other physical parameters is needed \citep{Wolff2011}. \textit{The authors would like to leave the reader with a thought: We have successfully shown that the basic building block of life is just a game of snakes/chutes (buckling) and ladders (pearling).}

\section*{Acknowledgments}
The authors would like to acknowledge support from National Science Foundation (grant 2147559-CBET).  

\section*{Declaration of Interests}
The authors report no conflict of interest.

\section{Appendix}
\subsection{Differential geometry basics}\label{appA}
Let us consider a cylindrical tube with  coordinates given by:
\begin{equation}
\textbf{x} = \left[a(z,\phi)\cos\phi,a(z,\phi)\sin\phi,z \right]
\end{equation}
Here, the tube radius $r = a(z,\phi)$ is written as follows:

\begin{equation}
    a(z,\phi) = 1 + \epsilon f(z,\phi) + \epsilon^2 g, \qquad \epsilon \ll 1
\end{equation}
where $f(z,\phi)$ is a small, spatially varying perturbation, and $g$ is a constant that ensures conservation of volume to $O(\epsilon^2)$.

Let us define the tangent vectors $\boldsymbol{x}_{\phi} = \frac{\partial \boldsymbol{x}}{\partial \phi}$ and $\boldsymbol{x}_{z} = \frac{\partial \boldsymbol{x}}{\partial z}$, as well as the normal vector $\boldsymbol{n} = \frac{ \boldsymbol{x}_{\phi} \times \boldsymbol{x}_z}{|\boldsymbol{x}_{\phi} \times \boldsymbol{x}_z|}$.  The double derivatives are also defined as $\boldsymbol{x}_{\phi \phi} = \frac{\partial^2 \boldsymbol{x}}{\partial \phi \partial \phi}$, $\boldsymbol{x}_{\phi z} = \frac{\partial^2 \boldsymbol{x}}{\partial \phi \partial z}$, and $\boldsymbol{x}_{zz} = \frac{\partial^2 \boldsymbol{x}}{\partial z \partial z}$.  After performing these operations, we evaluate the metric tensor $\boldsymbol{g}$ and curvature tensor $\boldsymbol{B}$ below:

\begin{equation}
\boldsymbol{g} = \begin{bmatrix}
                \boldsymbol{x}_{\phi} \cdot \boldsymbol{x}_{\phi} & \boldsymbol{x}_{\phi} \cdot \boldsymbol{x}_{z} \\
                \boldsymbol{x}_{z} \cdot \boldsymbol{x}_{\phi} & \boldsymbol{x}_{z} \cdot \boldsymbol{x}_{z} \\
                
\end{bmatrix} \qquad
\boldsymbol{B} = -\begin{bmatrix}
                \boldsymbol{n} \cdot \boldsymbol{x}_{\phi \phi} & \boldsymbol{n} \cdot \boldsymbol{x}_{\phi z} \\
                \boldsymbol{n} \cdot \boldsymbol{x}_{z \phi} & \boldsymbol{n} \cdot \boldsymbol{x}_{zz} \\
                
\end{bmatrix} \qquad
\end{equation}

The mean and Gaussian curvatures are obtained by the following formulas:

\begin{equation}
    2H = \text{Tr}(\boldsymbol{g}^{-1} \cdot \boldsymbol{B}) \qquad     K = \text{det}(\boldsymbol{g}^{-1} \cdot \boldsymbol{B}) 
\end{equation}
while the area element for the surface is given below, where $J$ is the surface Jacobian:

\begin{equation} \label{eq:surf_Jac_defn}
    dS = J d\phi dz; \qquad J = \sqrt{\text{det}(\boldsymbol{g})}
\end{equation}

Up to $O(\epsilon^2)$, the mean curvature and surface Jacobian are:

\begin{equation}
    2H = 1 - \epsilon \left( f + \frac{\partial^2 f}{\partial \phi^2} + \frac{\partial^2 f}{\partial z^2} \right) + \epsilon^2 \left[ f^2 - \frac{1}{2} \left( \frac{\partial f}{\partial z} \right)^2 + \frac{1}{2} \left( \frac{\partial f}{\partial \phi} \right)^2 + 2f \frac{\partial^2 f}{\partial \phi^2} - g \right]
\end{equation}

\begin{equation} \label{eq:surf_Jac}
    J  = 1 + \epsilon f + \epsilon^2 \left[ g + \frac{1}{2} \left( \frac{\partial f}{\partial z} \right)^2 + \frac{1}{2} \left( \frac{\partial f}{\partial \phi} \right)^2 \right]
\end{equation}

Up to $O(\epsilon)$, the Gaussian curvature is:

\begin{equation}
    K = -\epsilon \frac{\partial^2 f}{\partial z^2}
\end{equation}


\subsection{Rationale behind dimensionless numbers}\label{app:dimensionlessnumbers}

In this section, we try to clear the air about multiple dimensionless parameters using previous studies \citep{CamleyBrown2014,safran2018statistical}. According to the mentioned studies, the three experimentally measurable parameters that determine the dimensionless variables are the equilibrium concentration split ($\phi_{0}$), line tension ($\xi^{line}$), and interface width ($\varepsilon^{width}$). These dependencies are listed in equations \ref{eq:linetension_equation} and \ref{eq:width_equation}.

Moreover, 

\begin{equation}
    \phi_{0} = \sqrt{\frac{-b}{a}}
\end{equation}
These equations give us 

\begin{equation}
    \gamma^{2}a = \frac{9\phi_{0}^{4}(\xi^{line})^{2}}{8}
\end{equation}
and
\begin{equation}
    \gamma^{2} = \frac{a (\varepsilon^{width})^{2}}{2}
\end{equation}
This gives us:
\begin{equation}
    a = \frac{3\phi_{0}^{2}(\xi^{line})}{2\varepsilon^{width}}
\end{equation}

\begin{equation}
     \gamma = \sqrt{\frac{3\phi_{0}^{2}\xi^{line}\varepsilon^{width}}{4}}
\end{equation}

Using these equations and the definition of Cahn number, 
\begin{equation}
    Cn = \frac{\gamma}{R\sqrt{\zeta_{0}}}
\end{equation}

Assuming that $\zeta_{0}\approx |a|$, we get:

\begin{equation}
    Cn = \frac{\varepsilon^{width}}{\sqrt{2}R}
\end{equation}

\subsection{Coefficients for axisymmetric modes}\label{appsec:axisymmetric}

The linear equations in Eq (\ref{eq:matrix_eq}) admit an analytical solution for axisymmetric modes ($n=0$).  We obtain:
\begin{equation}
    A^{in}_{k0} = \frac{\Dot{r}_{k0}(k^{2}+1)I_{1}}{k\Psi}; \quad B^{in}_{k0} = 0; \quad C^{in}_{k0} = -\frac{\Dot{r}_{k0}(kI_{0} - I_{1})}{k\Psi}
\end{equation}

\begin{equation}
    A^{out}_{k0} = -\frac{\Dot{r}_{k0}(k^{2}+1)K_{1}}{k\Xi}; \quad B^{out}_{k0} = 0; \quad C^{out}_{k0} = -\frac{\Dot{r}_{k0}(kK_{0} + K_{1})}{k\Xi}
\end{equation}
where $\Psi = I_{1}^{2}k^{2}-I_{0}^{2}k^{2} + 2I_{0}I_{1}k$ and $\Xi = K_{1}^{2}k^{2}-K_{0}^{2}k^{2} - 2K_{0}K_{1}k$.

These equations give rise to:

\begin{equation}
    \Lambda_{k0} = 2(k^{2}+1)\left[\frac{K_{1}^{2}}{\Xi} - \lambda \frac{I_{1}^{2}}{\Psi}\right]
\end{equation}



\subsection{Dispersion relationship, low Peclet number limit ($Pe \ll 1$)}\label{App:LowPe}

When $Pe \ll 1$, the coarsening time is much smaller than the bending time scale ($t_{coarsening}\ll t_{bending}$).   In this case, a psuedo-steady approximation can be applied where the vesicle at any instance of time has a fixed, inhomogeneous phospholipid distribution on the surface.  Mathematically, the term $F_{kn}$ in (\ref{eq:final_eq}) is zero, which yields the concentration distribution $q_{kn} = -\frac{M_{kn}}{V_{kn}} r_{kn}$.  Since $\Lambda_{kn} \dot{r}_{kn} = L_{kn} r_{kn} + M_{kn} q_{kn}$, one obtains the dispersion relation:

\begin{equation}\label{eq:Dynamic_Eqn_lowPe}
\dot{r}_{kn} = \frac{r_{kn}}{\Lambda_{kn}}\left[ L_{kn} - \frac{M_{kn}^2}{V_{kn}} \right]
\end{equation}

To find the marginal wavenumber at which the growth rate is zero, we equate the term in brackets in \eqref{eq:Dynamic_Eqn_lowPe} to zero, which yields:

\begin{equation}
    \Gamma(n^{2}+k^{2}-1)+3/2 + 2k^{2}+(n^{2}+k^{2})(n^{2}+k^{2}-5/2) - \frac{\alpha Cn^{2}\beta^{2}(n^{2}+k^{2}-1)^{2}}{Cn^{2}(n^{2}+k^{2})+\tilde{a}} = 0
\end{equation}

We can obtain the marginal wavenumber for each mode $n=0,1,2..$. If we ignore the bending inhomogeneity and line tension by setting $Cn = \beta = 0$, this recovers the single-component vesicle result by \cite{boedec_jaeger_leonetti_2014}.  Lastly, if we consider the case where $\tilde{a}=-1$ (see Table \ref{tbl:Dimensionless_parameter_range}), we find that when $Cn^{2} > 1/k^{2}$, the growth rate is greater for a multicomponent vesicle compared to a single-component vesicle at the same surface tension conditions.

\subsection{Most unstable wavenumber dependence on memrbane tension}\label{app:wavenumber_Dependence}

\begin{figure}
\centering

\includegraphics[width=0.7\textwidth]{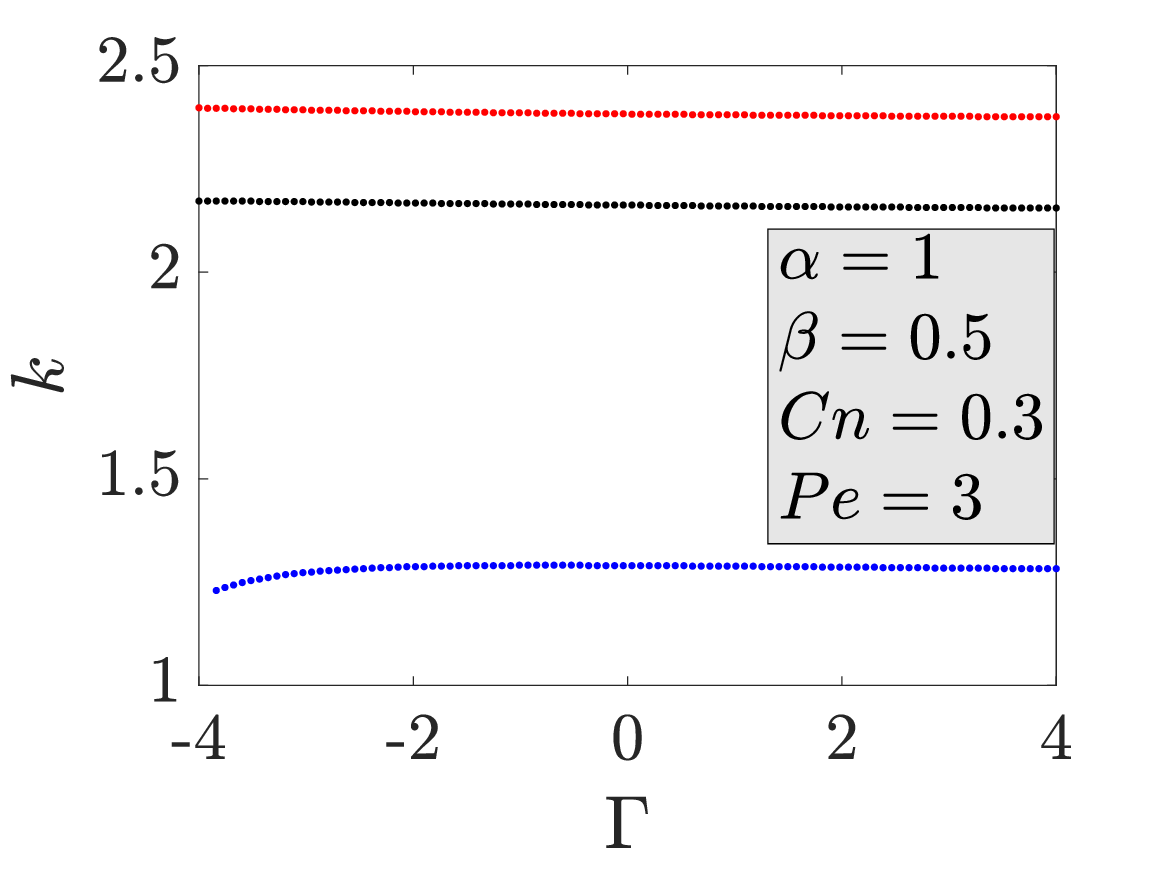}

\caption{Most unstable wave numbers with respect to the isotropic membrane tension $\Gamma$ for single-component vesicles. The red dots represent $n=0$ pearling modes, black dots represent $n=1$ buckling modes, and blue dots represent $n=2$ wrinkling modes. In the plot, $\lambda =1$.}
\label{fig:Most_Unstable_wavenumber_wrt_Gamma_Multi}
\end{figure}

In figure \ref{fig:Most_Unstable_wavenumber_wrt_Gamma_Multi}, we inspect the variation of the most unstable wavenumber with respect to the isotropic membrane tension. We can see that the most unstable wavenumber follows a gradual change with the membrane tension $\Gamma$. The pearling ($n=0$) and buckling modes ($n=1$) show a gradual drop in the wavenumber as the membrane tension increases, while the wrinkling wavenumbers show a slight increase with an increase in the membrane tension. The wavenumber behaviour for $n=0,n=1$ is consistent with the trend for single-component vesicles, albeit a much smaller decline in the magnitude. This indicates that the compressive membrane tension drives a shorter wavelength instability as compared to positive tension values.

\subsection{Derivation of energy change expressions for a deformed cylindrical vesicle}\label{app:energy}

We calculate the energy change of a perturbed vesicle from its unperturbed state, i.e., $\Delta E = E - E[r_{kn},q_{kn} = 0]$. Without loss in generality, let write the radius and concentration of the perturbed vesicle as:

\begin{equation}\label{radius}
        r = 1 + \epsilon r_{kn} \cos{(kz+n\phi)} - \frac{1}{4}\epsilon^{2} r_{kn}^2  
\end{equation}

\begin{equation}\label{radius}
        q = \epsilon q_{kn} \cos{(kz+n\phi)} - \frac{1}{2} \epsilon^2 q_{kn} r_{kn}  
\end{equation}

The $\epsilon^2$ term is added to the radius so that to $O(\epsilon^2)$, the volume of the vesicle $V = \int \int \frac{1}{2}r^2 d\phi dz$ is equal to its original volume $V_0 = \frac{1}{2} \int \int d\phi dz$.  The $\epsilon^2$ term is added to the concentration field so that the order parameter is conserved to $O(\epsilon^2)$ -- i.e.,  $\int q dS = 0$. 
 Using Eqs. (\ref{eq:surf_Jac_defn}) and (\ref{eq:surf_Jac}), the surface element along the vesicle is given by $dS = J d\phi dz$, with the surface Jacobian given by
 \begin{equation}
    J = 1 + \epsilon r_{kn} \cos{(kz + n\phi)} + \frac{\epsilon^2 r_{kn}^2}{4}\left[n^2 + k^2 - 1 - (n^2 + k^2)\cos{(2kz + 2n\phi)}\right]
\end{equation}

The energy contribution from surface tension is
\begin{equation}
    E_{\sigma}=\Gamma\int{dS}
\end{equation}

We perform the above integration, noting that only the zero-th order harmonics (i.e., constant terms) contribute to the integral.  This yields an energy change per unit length
\begin{equation}\label{tension_energy}
    \Delta E_{\sigma} = \frac{\Gamma\pi\epsilon^2 r_{kn}^2}{2}(n^2 + k^2 -1)
\end{equation}

The energy contribution from the phase behaviour given by the Landau-Ginzberg model.  In dimensionless form, it is:
\begin{equation}
    E_p = \frac{1}{Cn^2 \alpha}\int \frac{\tilde{a}}{2}|q|^2 + \frac{\tilde{b}}{4}|q|^4 + \frac{Cn^2}{2}|\nabla_s q|^2 dS
\end{equation}
We drop the middle term since it is $O(\epsilon^4)$ while 
\begin{equation}
    |\nabla_s q|^2 = \frac{(n^2 + k^2)\epsilon^2 q_{kn}^2}{2}[1-\cos{(2kn + 2n\phi)}]
\end{equation}
The energy change per unit length in this case is:
\begin{equation}\label{phase_energy}
    \Delta E_p = \frac{\pi \epsilon^2 q_{kn}^2}{2 \alpha Cn^2}\left[\tilde{a} + Cn^2 (n^2 + k^2)\right]
\end{equation}
The Canham-Helfrich bending energy is given by
\begin{equation}\label{CH_energy}
    E_b = \int 2(1+\beta q)|H|^2 dS = \int 2|H|^2 dS + \int 2\beta q |H|^2 dS
\end{equation}
The first integral in (\ref{CH_energy}) is same as that for a single-component cylindrical vesicle \citep{NarsimhanThesis} while the second integral gives a coupled energy term. The expressions are:
\begin{equation}\label{bending_energy}
    \Delta E_b = \frac{\pi \epsilon^2 {r_{kn}}^2}{2}\left(2k^{2} +{(k^{2}+n^{2})\left(k^{2}+n^{2}-\frac{5}{2}\right) + \frac{3}{2}}\right) + \beta \pi\epsilon^{2} r_{kn} q_{kn} (k^{2}+n^{2}-1)
\end{equation}

Lastly, we make a comment on the total change in free energy.  If we examine Eqs. (\ref{tension_energy}), (\ref{phase_energy}), and (\ref{bending_energy}), we see that the the total change in energy takes a quadratic form $\Delta E_{tot} = \frac{1}{2} \boldsymbol{y}^T \cdot \boldsymbol{E} \cdot{y}$, where $\boldsymbol{y} = \epsilon [r_{kn}, q_{kn}]^T$ and $\boldsymbol{E}$ is:

\begin{equation}
\boldsymbol{E} = \frac{1}{\epsilon^2} \begin{bmatrix}
                \frac{\partial^2 \Delta E_{tot}}{\partial r_{kn} \partial r_{kn}}  & \frac{\partial^2 \Delta E_{tot}}{\partial r_{kn} \partial q_{kn}} \\
                \frac{\partial^2 \Delta E_{tot}}{\partial q_{kn} \partial r_{kn}} & \frac{\partial^2 \Delta E_{tot}}{\partial q_{kn} \partial q_{kn}}
            \end{bmatrix} = 
            \pi \begin{bmatrix}
                L_{kn}  & M_{kn} \\
                M_{kn} & V_{kn}
            \end{bmatrix}
\end{equation}

Thus, the matrices $L_{kn}$, $M_{kn}$, and $V_{kn}$ in the linear stability analysis are related to the second variation in the free energy.  The quantity, $\frac{1}{\pi} \frac{\partial \Delta E_{tot}}{\partial \epsilon r_{kn}}$ gives the linearized, normal tractions on the interface (see Eq. (\ref{eqn:traction_balance}), while the $\frac{1}{\pi} \frac{\partial \Delta E_{tot}}{\partial \epsilon q_{kn}}$ gives the chemical potential on the interface (see Eq. \ref{eq:chemical_pot}).  Thus, the energy analysis is consistent with the linear stability analysis, although the energy analysis cannot give information on the time scale of instability or the most dangerous wavenumber.

\bibliographystyle{jfm}

\end{document}